\documentclass[graybox]{svmult}

\usepackage{mathptmx}       
\usepackage{helvet}         
\usepackage{courier}        
\usepackage{type1cm}        
\usepackage{makeidx}         
\usepackage{graphicx}        
\usepackage{multicol}        
\usepackage[bottom]{footmisc}

\usepackage{latexsym}
\usepackage{amsfonts}
\usepackage{amssymb}
\usepackage{amsmath}

\makeindex             

\begin{document}

\title*{AdS/CFT on the brane}
\author{Jiro Soda}
\institute{
Jiro Soda \at Department of Physics, Kyoto University, Kyoto 606-8501, Japan\\
 \email{jiro@tap.scphys.kyoto-u.ac.jp}
}
%
%
\maketitle

\abstract*{ }

\abstract{It is widely recognized that the AdS/CFT correspondence is a useful tool to
study strongly coupled field theories. On the other hand, Randall-Sundrum (RS)
 braneworld models have been actively discussed as a novel cosmological framework. 
 Interestingly, the geometrical set up of braneworlds is quite similar to
 that in the AdS/CFT correspondence. Hence, it is legitimate to seek a precise
 relation between these two different frameworks. 
 In this lecture, I will explain how the AdS/CFT correspondence is
 related to the RS braneworld models. 
There are two different versions of RS braneworlds,
namely, the single-brane model and the two-brane model.
 In the case of the single-brane model, we reveal the relation between
   the geometrical and the AdS/CFT correspondence 
   approach using the gradient expansion method.
It turns out that the high energy and the Weyl term corrections found in the geometrical
 approach correspond to the CFT matter correction found in the 
    AdS/CFT correspondence approach.   In the case of two-brane system, 
    we also show that the AdS/CFT correspondence plays
    an important role in the sense that the low energy effective field theory
    can be described by the conformally coupled scalar-tensor theory
     where the radion plays the role of the scalar field.
We also discuss dilatonic braneworld models from the point of view
of the AdS/CFT correspondence.
 }

\section{Introduction}

It is believed that string theory is a candidate of the unified theory of everything.
  Remarkably, string theory can be consistently formulated only 
  in 10 dimensions~\cite{sodPolchinski}. 
  This fact requires a mechanism to fill the gap between our real world and
 the higher dimensions.  Conventionally, the extra dimensions are considered
to be compactified to a small compact space of the order of the Planck scale.
 However, recent developments of superstring theory invented 
 a new idea, the so-called braneworld\index{braneworld} 
 where matter resides on the hypersurface in higher dimensional 
 spacetime~\cite{sodHorava:1995qa,sodHorava:1996ma,sodLukas:1998yy,sodAntoniadis:1998ig,sodArkaniHamed:1998nn}
 ( see also earlier independent works \cite{sodAkama:1982jy,sodRubakov:1983bb}).
  This hypersurface is called (mem)brane.
 This idea originates from D-brane solutions in string theory.
 Interestingly, the D-brane solution also gives rise to the 
 AdS/CFT\index{AdS/CFT} correspondence
 which claims that classical gravity in an anti-de Sitter(AdS) spacetime is equivalent to
 a strongly coupled conformally invariant field theory (CFT). 
 Since the origin is the same, braneworlds and the AdS/CFT correspondence
 may be related to each other. In particular, Randall and Sundrum (RS)
 braneworld models~\cite{sodRS1,sodRS2} have a similar geometrical setup to
 that in the AdS/CFT correspondence. Hence, 
  in this lecture, I will try to reveal relations between the AdS/CFT 
  correspondence and RS braneworld models~\cite{sodKanno:2004ns}. 
 
 The method we will use is the gradient expansion method\index{gradient expansion method}. 
 Physically, it is a low energy expansion method.
 Historically, the method has been used in the cosmological 
 context~\cite{sodtomita,sodSalopek:1992kk,sodComer,sodSII}. 
 In particular, it is known to be useful for analyzing the evolution of cosmological
 perturbations during inflation. Since the AdS spacetime can be regarded
 as the inflating universe in the spatial direction, the gradient
 expansion method is also expected to be useful in the AdS spacetime. 
 First, by utilizing the gradient expansion method, we approximately
 solve Einstein equations in the bulk. 
 Then, the junction conditions\index{junction conditions} at the brane
  give the effective equations of motion on the brane.  
 Thus, we can understand the low energy physics in the
 braneworld. The similar but slightly different method is 
 also used in the AdS/CFT correspondence.
 We will identify a concrete relation between the geometrical
 and the AdS/CFT correspondence approach by detailed comparison. 
 The difference shows up when we consider two brane systems. 
 Indeed, we do not have the conventional
 AdS/CFT correspondence for the two-brane system. Instead, we have
 a conformally coupled radion on the brane which reflects the conformal
 symmetry of the theory. This observation is useful for understanding 
 why brane inflation suffers from the eta problem. It is apparent that
 the gradient expansion method can be applicable to various braneworld models.
 Moreover, as we will see, the gradient expansion method
 provides a unified view of branworlds and a useful tool to
 make cosmological predictions. 

 The organization of this lecture is as follows:
 In section 2, we introduce RS models and derive the effective Friedman equation
  on the brane. Here, two important corrections, i.e., the dark radiation and
  the high energy corrections, are identified. 
  This sets our starting point.
  In section 3, we review two different views from the brane, namely,
  the AdS/CFT correspondence and the geometrical holography.
  These two frameworks give us a complementary picture of the braneworlds. 
  In section 4, we present key questions to make our concerns manifest.
  In section 5, we review the gradient expansion method.
  In sections 6 and 7, we apply the gradient expansion method to the single-brane model
  and to the two-brane model, respectively. We obtain the effective
  theory for both cases. 
  In section 8, we give answers to the key questions.
  This completes explanation of relations between the AdS/CFT correspondence
  and RS braneworld models.  
  In section 9, we extend our analysis to models with a bulk scalar field.
  Since the presence of bulk fields would break the conformal invariance,
  it is interesting to consider dilatonic braneworlds in conjunction with
  the AdS/CFT correspondence.
  The final section is devoted to the conclusion.

\section{Braneworlds in AdS spacetime}
\setcounter{equation}0

In this section, we will introduce RS braneworld models. 
We will derive the effective Friedmann equation and identify the effects of
extra-dimensions. In this lecture, we will concentrate on
cosmology although we can apply the results to black hole physics.

\subsection{RS Models}

  Randall and Sundrum  proposed a simple model
   where a four-dimensional brane with the tension $\sigma$
   is embedded in the five-dimensional asymptotically anti-de Sitter (AdS) bulk
    with a curvature scale $\ell$.
   This single-brane model\index{single-brane model} 
   is described by the action~\cite{sodRS2}
\begin{eqnarray}
S&=&{1\over 2\kappa^2}\int d^5 x \sqrt{-g}
	\left( 
	{\cal R}+{12\over \ell^2}
	\right)  -\sigma \int d^4 x \sqrt{-h}
	+\int d^4 x \sqrt{-h} {\cal L}_{\rm matter} \ ,
\end{eqnarray}
where ${\cal R}$ and $\kappa^2  $  are the scalar curvature and 
gravitational constant in five-dimensions, respectively. We  impose 
 $Z_2$ symmetry on this spacetime, with the brane at
 the fixed point. The matter ${\cal L}_{\rm matter}$ is 
 confined to the brane. Throughout this lecture,
 $h_{\mu\nu}$ represents the induced metric on the brane.  
 Remarkably, the internal dimension is non-compact in this model.
 Hence, we do not have to care about the stability problem. 
 The basic equations consist of the equations of motion in the bulk
 and junction conditions at the brane position
  due to the presence of the brane.
Alternatively, the basic equations can be regarded as the 5-dimensional
Einstein equations with singular sources. 
Let us recall the 4-dimensional components of 5-dimensional Einstein tensor
can be expressed by
\begin{eqnarray}
\overset{(5)}{G}{}_{\mu\nu}&=& 
	\overset{(4)}{G}{}_{\mu\nu} 
       + {\cal L}_{n^\mu} \left[ K_{\mu\nu} - g_{\mu\nu} K \right] 
        + \cdots \ ,
\end{eqnarray}
where ${\cal L}_{n^\mu}$ denotes the Lie derivative along the 
  unit normal vector to the brane, $n^\mu$. 
 Here, we defined the extrinsic curvature by
\begin{eqnarray}
   K_{\mu\nu} = - \left( \delta_\mu^\rho - n_\mu n^\rho \right)
   \nabla_\rho n_\nu \ .
\end{eqnarray}
By integrating Einstein equations along the normal to the brane,
we obtain the jump of the extrinsic curvature
$ (K^+_{\mu\nu}- g_{\mu\nu} K^+ ) - (K^-_{\mu\nu}- g_{\mu\nu} K^- )$
from the left hand side and the total energy momentum tensor on the brane
 from the right hand side due to the delta function sources. 
Thus, taking into account the $Z_2$ symmetry $K_{\mu\nu}\equiv
K^+_{\mu\nu}=-K^-_{\mu\nu}$,
 we obtain the junction conditions 
\begin{eqnarray}
\left[K^\mu{}_\nu-\delta^\mu_\nu K\right]\Big|_{\rm at\ the\ brane} 
    = {\kappa^2 \over 2}\left(-\sigma\delta^\mu_\nu 
	+T^\mu{}_\nu \right)     \ .
\end{eqnarray}
Here, $T_{\mu\nu}$ represents the 
energy-momentum tensor of the matter. 
 
 Originally, they proposed the two-brane model\index{two-brane model}
  as a possible solution
 of the hierarchy problem~\cite{sodRS1}.
  The action reads
\begin{eqnarray}
S&=&{1\over 2\kappa^2}\int d^5 x \sqrt{-g} 
	\left( {\cal R} 
	+{12 \over \ell^2}
	\right) 	\nonumber \\
	&&\quad
	-\sum_{i=\oplus ,\ominus} \sigma_i \int d^4 x \sqrt{-h_i } 
+\sum_{i=\oplus ,\ominus}
	\int d^4 x \sqrt{-h} {\cal L}_{\rm matter}^i
        \ , \label{action-5d}
\end{eqnarray}
where $\oplus$ and $\ominus$ represent the positive and the negative
 tension branes, respectively. 
In principle, one can consider multiple-branes although they are not 
discussed in this lecture.

\subsection{Cosmology}

 The homogeneous cosmology of the single-brane model
 is easy to analyze~\cite{sodBrax:2004xh}. Because of the Birkoff theorem due to the symmetry on the brane, 
 it is sufficient to consider AdS black hole spacetime:
\begin{eqnarray}
  ds^2 = - h(r) dt^2 + \frac{dr^2}{h(r)} 
  + r^2 \left[ d\chi^2 
  + f_k^2 (\chi) \left( d\theta^2 +\sin^2 \theta d\phi^2 \right) \right]
     \ ,
\end{eqnarray}
where 
\begin{eqnarray}
 f_k = \left\{ 
 \begin{array}{cc}
 \sin\chi & {\rm for} \ k=1 \\
 \chi & {\rm for}\  k=0 \\
 \sinh\chi & {\rm for}\  k=-1
 \end{array}
 \right.
\end{eqnarray}
and
\begin{eqnarray}
  h(r) = k -\frac{M}{r^2} + \frac{r^2}{\ell^2} \ .
\end{eqnarray}
Note that $M$ is the mass of the black hole, $k$ is the curvature of the horizon
and $\ell$ is the AdS curvature radius.
Suppose the brane is moving in this spacetime with a trajectory
$t=t(\tau) \ , r=a(\tau)$, where $\tau$ is a proper time of the brane
(see Fig.1).
The induced metric on the brane becomes
\begin{eqnarray}
  ds_{ind}^2 = -d\tau^2 + a^2 (\tau) \left[ d\chi^2 
  + f_k^2 (\chi) \left( d\theta^2 +\sin^2 \theta d\phi^2 \right) \right]
  \ .
\end{eqnarray}
This is nothing but the  Friedman-Robertson-Walker spacetime where
$a$ is the scale factor. 
The motion of the brane cannot be arbitrary. It is constrained
by the junction condition:
\begin{eqnarray}
   K^\chi{}_\chi 
   =-\frac{\kappa^2 \sigma}{6}
    - \frac{\kappa^2}{2} \left[ T^\chi{}_\chi - \frac{1}{3} T \right] \ , 
    \label{sodjunctionchi}
\end{eqnarray}
where $K^\chi{}_\chi$, $T^\chi{}_\chi$, $T$ are a $\chi\chi$ component of
 the extrinsic curvature, a $\chi\chi$ component and the trace part of
the energy momentum tensor of matter on the brane, respectively. 
\begin{figure}[h]
\centerline{\includegraphics[width = 6cm, height = 7cm]{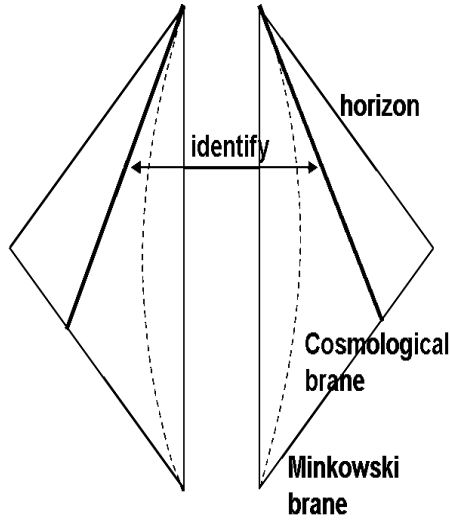}}
\caption{The Minkowski brane represented by the dotted line
is a static brane in the Poincare coordinate system
of the AdS spacetime. While, the cosmological brane represented by
the thick line is moving in the bulk. 
The motion of the brane induces the expansion of the brane universe.
The Cauchy horizon of AdS spacetime corresponds to the big-bang singularity.
 In the case of AdS-Schwarzschild spacetime, 
 the horizon should be the past horizon of the black hole. The big bang is 
 located beyound the horizon.}
\end{figure}
From the normalization condition $n^\mu n_\mu =1$
of the unit normal vector $n^\mu = (- \dot{a}, \dot{t} )$, we obtain
\begin{eqnarray}
   \dot{t} = \sqrt{\frac{1}{h(a)}+\frac{\dot{a}^2}{h^2(a)}} \ .
   \label{sodt}
\end{eqnarray}
Here, the dot is a derivative with respect to the proper time $\tau$. 
Now, one can calculate $K_{\chi\chi}$ as
\begin{eqnarray}
  K_{\chi\chi} = - \nabla_\chi n_\chi 
  = - \partial_\chi n_\chi + \Gamma^r_{\chi\chi} 
  = - \frac{1}{2} n^r \frac{\partial}{\partial \chi} g_{\chi\chi} \ . 
  \label{sodkchi}
\end{eqnarray}
Hence, from Eqs.(\ref{sodjunctionchi}), (\ref{sodt}) and (\ref{sodkchi}), 
we have an equation
\begin{eqnarray}
  \frac{1}{a}\sqrt{h(a) + \dot{a}^2 } = \frac{\kappa^2}{6} \left(\sigma+\rho\right) \ .
\end{eqnarray}
Thus, we get
\begin{eqnarray}
\frac{k}{a^2} -\frac{M}{a^4} + \ell^2 + \frac{\dot{a}^2}{a^2} 
= \frac{\kappa^4}{36} \left( \sigma+\rho\right)^2 \ ,
\end{eqnarray}
or
\begin{eqnarray}
   H^2 = \frac{\kappa^4 \sigma^2}{36} - \frac{1}{\ell^2} 
   +\frac{\kappa^4 \sigma }{18} \rho - \frac{k}{a^2} + \frac{M}{a^4}
   + \frac{\kappa^4}{36} \rho^2 \ .
\end{eqnarray}
By setting $\kappa^2 \sigma \ell = 6$, we finally derived
  the effective Friedmann equation
 as~\cite{sodBinetruy:1999hy,sodKraus:1999it,sodFlanagan:1999cu,sodIda:1999ui,sodKaloper:1999sm}
\begin{eqnarray}
   H^2 = - \frac{k}{a^2}+{\kappa^2 \over 3\ell} \rho  
            + \frac{M}{a^4} + \frac{\kappa^4}{36} \rho^2 \ ,
     \label{sodFRW}
\end{eqnarray}
where $H=\dot{a}/a$ is the Hubble parameter.
The Newton's constant can be identified as $8\pi G_N = \kappa^2 /\ell$. 
The curvature of the horizon $k$ corresponds spatial curvature of 
 the universe.  
 The black hole mass $ M$ is referred to as the 
 dark radiation\index{dark radiation}~\cite{sodMukohyama:1999qx} which
 is not real radiation fluid but a reflection of the bulk geometry.
 This effect exists even in the low energy regime.
The last term represents the high energy effect of the 
braneworld~\cite{sodMaartens:2003tw}.  

 As to the two-brane model, the same effective Friedmann equation (\ref{sodFRW}) can be
 expected on each brane because the above equation (\ref{sodFRW})
 has been deduced without referring to the bulk equations of motion.     
 
Given this cosmological background, it is natural to investigate cosmological
 perturbation in the braneworld~\cite{sodMaartens:2004yc}. 
  In the case of the single-brane model,
  it is shown that the gravity in Minkowski brane is localized on the brane in spite of 
  the noncompact extra dimension. Consequently, it turned out that 
   the conventional linearized Einstein equation 
   approximately holds at  scales large compared with
   the curvature scale $\ell$. It should be stressed that this result
   can be attained by imposing the outgoing boundary conditions. 
It turns out that this is also true in the cosmological background~\cite{sodKoyama:2000cc}. 
  
   In the case of the two-brane model, 
Garriga and Tanaka analyzed linearized gravity
 and have shown that the gravity  on the brane 
behaves as the Brans-Dicke theory\index{Brans-Dicke theory} at low energy~\cite{sodGT}.
 Thus, the conventional linearized Einstein equations 
do not hold even on scales large compared with the curvature scale $\ell$ 
in the bulk. Charmousis et al. have clearly identified the Brans-Dicke field 
as the radion\index{radion} mode~\cite{sodCR}. 

In the end, we would like to know how nonlinear gravity in the braneworld
is deviated from the conventional Einstein gravity.
A partial answer will be given in the following sections.
  
\section{View from the brane}
 
 In the previous section, we have considered an isotropic and homogeneous universe
 and seen that the effective Friedmann equation
 on the brane can be regarded as the conventional Friedmann equation
 with two kind of corrections, i.e., the dark radiation and high-energy corrections.
 Here, we review two different approaches to extend the above result to
 more general cases. 
 
\subsection{AdS/CFT Correspondence }
 
 Let us start with the AdS/CFT correspondence~\cite{sodmalda,sodGubser:1998bc,sodWitten:1998qj}.
After solving the equations of motion in the bulk with the boundary
 value fixed and substituting the solution $g_{\rm cl}$ 
 into the 5-dimensional Einstein-Hilbert action $S_{\rm 5d}$,
 we obtain the effective action for the boundary field
  $h= g_{\rm cl}|_{\rm boundary}$.
 The statement of the AdS/CFT correspondence is that the resultant
 effective action can be equated with the partition functional 
 of some conformally invariant field theory (CFT), namely     
\begin{eqnarray}
   \exp \left[ i S_{\rm 5d} [ g_{\rm cl}] \right] 
  \approx < \exp \left[ i \int h {\cal O} \right] >_{\rm CFT} \ , 
\end{eqnarray}
where ${\cal O} $ is the field in CFT.
In the right hand side, $h$ should be interpreted as a source field.
This action must be defined at the AdS infinity where the conformal symmetry
 exists as the asymptotic symmetry. Hence, there exist infrared divergences
 which must be subtracted by counter terms. 
 Thus, the correct formula becomes
\begin{eqnarray}
   \exp \left[ i S_{\rm 5d} [ g_{\rm cl}] + i S_{\rm ct} \right] 
  = < \exp \left[ i \int h {\cal O} \right] >_{\rm CFT} \ , 
\end{eqnarray}
where we added the counter terms
\begin{eqnarray}
    S_{\rm ct} =  S_{\rm brane} - S_{\rm 4d} 
                     - [~R^2 {\rm terms}~]    \ ,
\end{eqnarray}
where $S_{\rm brane}$ and $S_{4d}$ are the brane action  and
the 4-dimensional Einstein-Hilbert action, respectively.
Here, the higher curvature terms $[R^2 {\rm terms}]$ should be 
understood symbolically.

 In the case of the braneworld, the brane acts as the cutoff.
 Therefore, there is no divergences in the above expressions.
 In other words, no counter term is necessary. We can regard the
 above relation as the definition of the ``cut off" CFT.
 Thus, we can freely rearrange the terms as follows 
\begin{eqnarray}
S_{\rm 5d}+S_{\rm brane}=S_{\rm 4d}+S_{\rm CFT}+[~R^2 {\rm terms}~] \ ,
\end{eqnarray}
where we have defined 
\begin{eqnarray}
 \exp iS_{\rm CFT} 
 \equiv < \exp \left[ i \int h {\cal O} \right] >_{\rm CFT}  \ .
\end{eqnarray}
This tells us that the brane models can be described as the conventional
 Einstein theory with the cutoff CFT and higher order curvature 
 terms~\cite{soddeHaro:2000wj,sodads1,sodads2,sodads3}.
In terms of the equations of motion, the AdS/CFT correspondence reads
\begin{equation}
\overset{(4)}{G}{}_{\mu\nu} 
	={\kappa^2\over\ell}\left( T_{\mu\nu}+T^{\rm CFT}_{\mu\nu} \right)
	+[~R^2 {\rm terms}~]   \ ,
      \label{sodCFTeq}
\end{equation}
where the $R^2$ terms represent the higher order curvature terms and 
 $T^{CFT}_{\mu\nu}$ denotes the energy-momentum tensor of the cutoff version 
 of  conformal field theory. 
When we apply this result to cosmology, we see CFT corresponds to
the dark radiation in the braneworld and the higher curvature terms 
can be reduced to the high-energy corrections.
 
\subsection{Geometrical Holography}

Here, let us review the geometrical approach~\cite{sodShiMaSa}.  
In the  Gaussian normal coordinate system: 
\begin{eqnarray}
ds^2=dy^2+g_{\mu\nu}(y,x^{\mu} )dx^{\mu}dx^{\nu}  \ ,
\label{sodGN}
\end{eqnarray}
we can write  the 5-dimensional Einstein tensor $\overset{(5)}{G}{}_{\mu\nu}$
in terms of the 4-dimensional Einstein tensor $\overset{(4)}{G}{}_{\mu\nu}$
 and the extrinsic curvature as 
\begin{eqnarray}
\overset{(5)}{G}{}_{\mu\nu}&=& 
	\overset{(4)}{G}{}_{\mu\nu} + K_{\mu\nu,y} - g_{\mu\nu} K_{,y}
	-KK_{\mu\nu}+2K_{\mu\lambda}K^{\lambda}{}_{\nu} \nonumber \\
	&&\qquad+{1\over 2}g_{\mu\nu} 
	\left(K^2+K^\alpha{}_\beta K^\beta{}_\alpha \right)  \nonumber\\
	&=&{6\over\ell^2}g_{\mu\nu}\ ,
	\label{sodGauss}
\end{eqnarray}
where  we have introduced the extrinsic curvature 
\begin{eqnarray}
K_{\mu\nu} = - {1\over 2} g_{\mu\nu ,y}  \ ,
\end{eqnarray}
and the last equality comes from the 5-dimensional Einstein equations.
On the other hand, the Weyl tensor in the bulk can be expressed as 
\begin{eqnarray}
  C_{y\mu y\nu} = K_{\mu\nu ,y} - g_{\mu\nu} K_{,y}
  + K_\mu{}^\lambda K_{\lambda\nu} 
  + g_{\mu\nu} K^{\alpha \beta} K_{\alpha\beta}
   - \frac{3}{\ell^2} g_{\mu\nu}  
   \label{sodWeyl}\ .
\end{eqnarray}
Now, one can eliminate $K_{\mu\nu ,y} - g_{\mu\nu} K_{,y}$ 
from (\ref{sodGauss}) using (\ref{sodWeyl})
and obtain
\begin{eqnarray}
  \overset{(4)}{G}{}_{\mu\nu}&=& 
	 - C_{y\mu y\nu}
	+KK_{\mu\nu}- K_{\mu\lambda}K^{\lambda}{}_{\nu} \nonumber \\
	&&\qquad - {1\over 2}g_{\mu\nu} 
	\left(K^2 - K^\alpha{}_\beta K^\beta{}_\alpha \right) 
      + {3\over\ell^2}g_{\mu\nu} \ .
      \label{sodpreSMS}
\end{eqnarray}

Taking into account the $Z_2$ symmetry,   we also obtain the junction 
 conditions 
\begin{eqnarray}
\left[K^\mu{}_\nu-\delta^\mu_\nu K\right]\Big|_{y=0} 
    = {\kappa^2 \over 2}\left(-\sigma\delta^\mu_\nu 
	+T^\mu{}_\nu \right)     \ .
\end{eqnarray}
Here, $T_{\mu\nu}$ represents the 
energy-momentum tensor of the matter. 
Evaluating Eq.~(\ref{sodpreSMS}) at the brane and substituting the 
junction condition 
into it, we have the ``effective" equations of motion 
\begin{eqnarray}
\overset{(4)}{G}{}_{\mu\nu} 
	={\kappa^2\over\ell}T_{\mu\nu}+\kappa^4\pi_{\mu\nu} 
	-E_{\mu\nu}
	\label{sodShiMaSa}
\end{eqnarray}
where we have defined the quadratic of the energy momentum tensor
\begin{eqnarray}                   
 \pi_{\mu\nu} =-{1\over 4}T_\mu{}^\lambda T_{\lambda\nu} 
  +{1\over 12}TT_{\mu\nu}
    + {1\over 8} g_{\mu\nu} \left( 
   T^{\alpha\beta}T_{\alpha\beta} -{1\over 3} T^2 \right) 
\end{eqnarray}
and the projection of Weyl tensor $C_{y\mu y \nu}$ onto the brane
\begin{eqnarray}
   E_{\mu\nu} &=& C_{y\mu y \nu} |_{y=0}  \nonumber \ .
\end{eqnarray}
Here, we assumed the relation
\begin{eqnarray}
\kappa^2 \sigma = \frac{6}{\ell}
\label{sodrelation}
\end{eqnarray}
so that the effective
 cosmological constant vanishes. 

Because of the traceless property of $E_{\mu\nu}$, when we consider
an isotropic and homogeneous universe, it is easy to show that
this gives the dark radiation component $\propto 1/a^4$. The existence of the
high-energy corrections $\propto \rho^2 $ is apparent in this approach.

The geometrical approach is useful to classify possible corrections
 to the conventional Einstein equations. One defect of this approach
 is the fact that the projected Weyl tensor $E_{\mu\nu}$ 
 can not be determined without solving the equations in the bulk. 
 
\section{Does AdS/CFT play any role in braneworld?}

To make our concerns explicit, we give a sequence of questions.
 We treat the single-brane model and two-brane model, separately.

\subsection{Single-brane model}

\noindent
{ \bf  Is the Einstein theory recovered even in the non-linear regime?}\\

In the case of the linear theory, it is known that the conventional
 Einstein theory is recovered at low energy. 
 On the other hand, the cosmological consideration suggests the deviation
 from the conventional Friedmann equation even in the low energy regime. 
 This is due to the dark radiation term. 
 Therefore, we need to clarify when the conventional Einstein theory
 can be recovered on the brane.  

\vskip 0.5cm
\noindent
{\bf How does the AdS/CFT come into the braneworld?}\\

 It was argued that the cutoff CFT comes into the braneworld.
 However, no one knows what is the cutoff CFT.
 It is a vague concept at least from the point of view of the classical
 gravity. 
  Moreover, it should be noted that the AdS/CFT  correspondence is a specific 
  conjecture. Indeed, originally, Maldacena conjectured that 
  the supergravity on $AdS_5 \times S^5$ is dual to the four-dimensional 
 ${\cal N}=4$ super Yang-Mills theory~\cite{sodmalda}. 
 Nevertheless, the AdS/CFT correspondence seems to
 be related to the brane world model as has been  demonstrated 
 by several people~\cite{sodads1,sodads2,sodads3,sodads4,sodads5,
 sodads6,sodads7,sodKiritsis:2005bm,sodTanaka:2004ig}.
 Hence, it is important to reveal the role of the AdS/CFT correspondence
 starting from the 5-dimensional general relativity.

\vskip 0.5cm
\noindent
{\bf How are the AdS/CFT and geometrical approach related?}\\

The geometrical approach gives the effective equations of motion (\ref{sodShiMaSa})
\begin{eqnarray*}
 \overset{(4)}{G}{}_{\mu\nu} = {\kappa^2 \over \ell} T_{\mu\nu} 
               + \kappa^4 \pi_{\mu\nu} 
                   - E_{\mu\nu}  \ .
\end{eqnarray*}
 On the other hand, the AdS/CFT correspondence yields 
 the other effective equations of motion (\ref{sodCFTeq})
\begin{equation}
   \overset{(4)}{G}{}_{\mu\nu} 
   = {\kappa^2 \over \ell} \left( T_{\mu\nu} + T^{\rm CFT}_{\mu\nu} \right)
                   +[~R^2 {\rm terms}~]   \ .\nonumber
\end{equation}
An apparent difference is remarkable. 

It is an interesting issue to 
 clarify how  these two descriptions are related.  
  Shiromizu and Ida tried to understand the AdS/CFT correspondence 
  from the  geometrical point of view~\cite{sodSI}.
   They argued that $\pi^\mu_{\ \mu}$ 
  corresponds to the trace anomaly of the cutoff CFT on the brane. 
  However, this  result is rather  paradoxical because there exists 
  no trace anomaly 
  in an odd dimensional  brane although  $\pi^\mu_{\ \mu}$ exists even in that 
  case.  Thus,  the more precise relation between the 
   geometrical and the AdS/CFT approaches should be given. 
  
   Moreover, since both the geometrical and AdS/CFT approaches seem 
  to have their own merit, 
  it would be beneficial to understand the mutual relationship.

\subsection{Two-brane model}

\noindent
{\bf How is the geometrical approach consistent with the Brans-Dicke
 picture?}\\

Irrespective of the existence of other branes, the geometric approach gives
 the same effective equations (\ref{sodShiMaSa}). 
The effect of the bulk geometry comes into
the brane world only through $E_{\mu\nu}$. 
In this picture, the two-brane system can be regarded as the Einstein
 theory with some corrections due to the Weyl tensor in the bulk.

On the other hand,  the linearized gravity  on the brane 
behaves as the Brans-Dicke 
theory  on scales large compared with the curvature scale $\ell$ 
in the bulk~\cite{sodGT}. 
 Therefore, the conventional linearized Einstein equations 
 do not hold at low energy. 

In the geometrical approach, no radion appears. While, from the linear
 analysis, it turns out that 
 the system can be described by the Brans-Dicke theory where
 the extra scalar field is nothing but the radion.  
How can we reconcile these seemingly incompatible pictures? 

\vskip 0.5cm
\noindent
{\bf What replaces  the AdS/CFT correspondence  
in the two-brane model?}\\

In the single-brane model, 
there are continuum Kaluza-Klein (KK)-spectrum around the zero mode.
 They induce the CFT matter in the 4-dimensional effective action.
 In the two-brane system, the spectrum become discrete and then a mass gap exists.
 Hence, we can not expect CFT matter on the brane, although
 KK-modes exist and  affect the physics on the brane. 
 So, it is  interesting to know if the AdS/CFT correspondence
  play a role in the two-brane system.

\section{Gradient Expansion Method}

 Our aim in this lecture is to show the gradient expansion
 method gives the answers to all of the questions presented in the 
 previous section. Here, we review the formalism developed by us~
\cite{sodKS1,sodKS2,sodKS3,sodKS4}.  

We use the Gaussian normal coordinate\index{Gaussian normal coordinate}
 system (\ref{sodGN}) to describe
the geometry of the brane world. 
Note that the brane is located at $y=0$ in this coordinate system. 
Decomposing the extrinsic curvature into the traceless part and the trace part
\begin{equation}
  K_{\mu\nu} 
    = \Sigma_{\mu\nu }+ {1\over 4} h_{\mu\nu} K  \ , \quad
    K = - {\partial \over \partial y}\log \sqrt{-g}    \  , 
\end{equation}
we obtain the basic equations which hold in the bulk;
\begin{eqnarray}
&&\Sigma^\mu{}_{\nu ,y}-K\Sigma^\mu{}_{\nu} 
	=-\left[
	\overset{(4)}{R}{}^\mu{}_\nu 
	-{1\over 4} \delta^\mu_\nu \overset{(4)}{R} 
        \right]
        \label{sodmunu-trc} \ ,     \\
&&{3\over 4}K^2-\Sigma^\alpha{}_{\beta}\Sigma^\beta{}_{\alpha} 
	=\left[~
	\overset{(4)}{R}
	~\right] 
	+{12\over\ell^2}
	\label{sodmunu-trclss} \ ,  \\
&&K_{, y}-{1\over 4}K^2-\Sigma^{\alpha\beta}\Sigma_{\alpha\beta} 
	=-{4\over\ell^2}
	\label{sodyy} \ ,  \\
&&\nabla_\lambda\Sigma_{\mu}{}^{\lambda}  
	-{3\over 4}\nabla_\mu K = 0
	\label{sodymu} \ ,
\end{eqnarray}
where $\overset{(4)}{R}{}^\mu{}_\nu$ 
is the curvature on the brane and $\nabla_\mu $ denotes the
 covariant derivative with respect to the metric $g_{\mu\nu}$.
 One also has the junction condition 
\begin{equation}
\left[
	K^\mu{}_\nu -\delta^\mu_\nu K
	\right]
	\Big|_{y=0} 
	={\kappa^2\over 2}
	\left(
	-\sigma\delta^\mu_\nu+T^\mu{}_\nu
	\right) \ .
\end{equation}
Recall that we are considering the $Z_2$ symmetric spacetime. 

The problem now is separated into two parts. First, we will solve
 the bulk equations of motion with the Dirichlet boundary condition
 at the brane, 
\begin{eqnarray}
 g_{\mu\nu} (y=0 ,x^\mu ) = h_{\mu\nu} (x^\mu ) \ .
 \label{sodBC}
\end{eqnarray}
 After that, the junction condition will be imposed at the brane.
 As it is the condition for the induced metric $h_{\mu\nu}$, it
 is naturally interpreted as the effective equations of motion
 for  gravity on the brane.
 
Along the normal coordinate $y$, the metric varies with a characteristic length
 scale $\ell$; $ g_{\mu\nu ,y} \sim g_{\mu\nu} /\ell$.  Denote the 
characteristic 
 length scale of the curvature fluctuation on the brane as $L$; then we have
  $ R \sim g_{\mu\nu} / L^2 $. For the reader's reference, let us take 
  $\ell=1$ mm, for example. Then, the relation (\ref{sodrelation}) give 
  the scale, 
  $\kappa^2 = (10^8 \ {\rm GeV})^{-3}$ and $\sigma = 1 \ {\rm TeV}^4$.  
  In this lecture, we will consider the low energy
  regime in the sense that the energy density of matter, $\rho$, 
  on the brane is smaller than the brane tension, \i.e., $\rho /\sigma \ll 1$. 
  In this regime, a simple dimensional analysis
\begin{equation}
  {\rho \over \sigma} 
  \sim {\ell {\kappa^2 \over \ell} \rho \over \kappa^2 \sigma}
    \sim \left({\ell\over L}\right)^2 \ll 1 
\end{equation}
implies that the curvature on the brane can be neglected compared with the
 extrinsic curvature at low energy. Here, we have used the relation 
 (\ref{sodrelation})
 and Einstein's equation on the brane, 
 $R\sim g_{\mu\nu}/L^2 \sim \kappa^2\rho/\ell$.
 Thus, the anti-Newtonian or gradient expansion method used in the cosmological
 context is applicable to our problem~\cite{sodtomita}.

At zeroth order,  we can neglect the curvature term. Then we have
\begin{eqnarray}
&&\overset{(0)}{\Sigma}{}^{\mu}{}_{\nu , y}
	-\overset{(0)}{K}\overset{(0)}{\Sigma}{}^{\mu}{}_{\nu} 
	=0
	\label{sod0:munu}     \ ,  \\
&&{3\over 4}\overset{(0)}{K}{}^{2} 
	-\overset{(0)}{\Sigma}{}^{\alpha}{}_{\beta} 
	\overset{(0)}{\Sigma}{}^{\beta}{}_{\alpha} 
	={12\over \ell^2}   \ ,  \\
&&\overset{(0)}{K}{}_{, y} -{1\over 4} \overset{(0)}{K}{}^{2} 
	-\overset{(0)}{\Sigma}{}^{\alpha \beta} 
	\overset{(0)}{\Sigma}{}_{\alpha \beta} 
	=-{4\over \ell^2}     \ ,   \\
&&\nabla_\lambda\overset{(0)}{\Sigma}{}^{\lambda}{}_{\mu }   
	-{3\over 4}\nabla_\mu \overset{(0)}{K} = 0
	\label{sod0:ymu} \ .
\end{eqnarray}
 Equation (\ref{sod0:munu}) can be readily integrated into
\begin{equation}
    \overset{(0)}{\Sigma}{}^{\mu}{}_{\nu} 
    = {C^\mu{}_{\nu} (x^\mu) \over \sqrt{-g} } 
                                 \ , \quad C^\mu{}_{\mu} =0  \ ,
\end{equation}
where $C^\mu{}_{\nu}$ is the constant of integration. 
 Equation (\ref{sod0:ymu}) also requires $C^\mu{}_{\nu |\mu} =0$.
 If it could exist, it would represent a radiation like fluid 
 on the brane and hence  a strongly anisotropic universe. 
 In fact, as we see soon, this
 term must vanish in order to satisfy the junction condition. 
 Therefore, we simply put $C^\mu{}_\nu =0$, hereafter. 
 Now, it is easy to solve the remaining equations. The result is
\begin{equation}
    \overset{(0)}{K} = {4\over \ell}    \ .
\end{equation}
Using the definition of the extrinsic curvature
\begin{equation}
     \overset{(0)}{K}{}_{\mu\nu} 
     = - {1\over 2} {\partial \over \partial y} 
                               \overset{(0)}{g}{}_{\mu\nu}     \   ,
\end{equation}
we get the zeroth order metric  as
\begin{equation}
 ds^2 = dy^2 +  b^2 (y) h_{\mu\nu}(x^\mu ) dx^\mu dx^\nu  \ , \quad
    b(y)  = e^{-2{y\over \ell}}    \ ,
\end{equation}
where  the tensor $h_{\mu\nu}$ is  the induced metric on the brane,
which conforms to the boundary condition (\ref{sodBC}). 

From the zeroth order solution, we obtain
\begin{equation}
   \left[ \overset{(0)}{K}{}^{\mu}{}_{\nu} 
   - \delta^\mu_\nu \overset{(0)}{K} \right] \Bigg|_{y=0}
    = -{3 \over \ell} \delta^\mu_\nu
    = - {\kappa^2 \over 2} \sigma \delta^\mu_\nu  \ .
\end{equation}
Then we get the well known relation 
\begin{eqnarray}
\kappa^2 \sigma = 6/\ell  \ . 
\end{eqnarray}
Here, we will assume that this relation holds exactly. 
 It is apparent that $C^\mu{}_{\nu}$ is not allowed to exist.

The iteration scheme consists in writing the metric $g_{\mu\nu}$
 as a sum of local tensors built out of the induced metric on the
 brane, the number of gradients increasing with the order. 
Hence, we will seek the metric as a perturbative series
\begin{eqnarray}
     g_{\mu\nu} (y,x^\mu ) &=&
  b^2 (y) \left[ h_{\mu\nu} (x^\mu)  + \overset{(1)}{g}{}_{\mu\nu} (y,x^\mu)
  + \overset{(2)}{g}{}_{\mu\nu} (y, x^\mu ) + \cdots  \right]  \ , 
\end{eqnarray}
where $b^2 (y) $ is extracted 
 and we put the Dirichlet boundary condition
\begin{eqnarray}  
     \overset{(i)}{g}{}_{\mu\nu} (y=0 ,x^\mu ) =  0    \ ,
\end{eqnarray}
so that $g_{\mu\nu} (y=0, x) =  h_{\mu\nu} (x)$ holds at the brane. 
Other quantities can be also expanded as
\begin{eqnarray}
K^\mu{}_{\nu}&=&{1\over\ell}
	\delta^{\mu}_{\nu}
        +\overset{(1)}{K}{}^{\mu}{}_{\nu}
	+\overset{(2)}{K}{}^{\mu}{}_{\nu}+\cdots  \nonumber\\
\Sigma^\mu{}_{\nu}
	&=&\qquad
	+\overset{(1)}{\Sigma}{}^{\mu}{}_{\nu}
	+\overset{(2)}{\Sigma}{}^{\mu}{}_{\nu} + \cdots          \ .
\end{eqnarray}
In our scheme,
 in contrast to the AdS/CFT correspondence where the Dirichlet boundary 
 condition is imposed at infinity, we impose it
  at the finite point $y=0$, the location of the brane. 
 Furthermore, we carefully consider the
  constants of integration, i.e., homogeneous solutions. These 
 homogeneous solutions are  ignored in the calculation of AdS/CFT 
 correspondence.  However, they play an important role in the
 braneworld.  Note that the scheme can be applicable to other 
 systems~\cite{sodShiromizu:2003dr,sodOnda:2003sj,sodShiromizu:2003pz,sodTakahashi:2004ss}.

\section{Single brane model (RS2)}

Now, we will apply the gradient expansion method to the single-brane models
and obtain the effective equations on the brane.

\subsection{Einstein Gravity at Lowest Order}

The next order solutions are obtained by taking into account the 
terms neglected at zeroth order. 
At  first order,  Eqs.~(\ref{sodmunu-trc}) - (\ref{sodymu}) become
\begin{eqnarray}
&&\overset{(1)}{\Sigma}{}^{\mu}{}_{\nu , y} 
	-{4\over\ell} \overset{(1)}{\Sigma}{}^{\mu}{}_{\nu} 
	=-\left[\overset{(4)}{R}{}^\mu{}_\nu 
	-{1\over 4} \delta^\mu_\nu\overset{(4)}{R}\right]^{(1)}
	\label{sod1:munu} \ , \\
&&{6 \over\ell} \overset{(1)}{K}  = \left[~ \overset{(4)}{R}
	~\right]^{(1)}   \  ,\\
&&\overset{(1)}{K}{}_{, y} -{2\over \ell} \overset{(1)}{K} = 0   \ ,\\
&&\overset{(1)}{\Sigma}{}_{\mu}{}^{\lambda}{}_{|\lambda}  
	-{3\over 4}\overset{(1)}{K}{}_{|\mu} = 0 \ .
\end{eqnarray}
where the superscript $(1)$ represents the order of the derivative expansion
 and a stroke $|$ denotes the covariant derivative with respect to
 the metric $h_{\mu\nu}$.
Here, $[\overset{(4)}{R}{}^\mu{}_\nu ]^{(1)} $ 
means that the curvature is approximated by
 taking the Ricci tensor of $b^2 h_{\mu\nu} $ in place of 
 $\overset{(4)}{R}{}^{\mu}{}_{\nu}$. 
 It is also convenient to write it in terms of the Ricci
  tensor of $h_{\mu\nu}$, denoted $R^\mu{}_\nu (h)$.
 
Substituting the zeroth order metric into $\overset{(4)}{R}$, we obtain
\begin{equation}
\overset{(1)}{K} = {\ell\over 6b^2} R(h)
\label{sod1:trc} \ .
\end{equation}
Hereafter, we omit the argument of the curvature for simplicity. 
Simple integration of Eq.~(\ref{sod1:munu}) also gives the traceless part
 of the extrinsic curvature as
\begin{equation}
\overset{(1)}{\Sigma}{}^{\mu}{}_{\nu}={\ell\over 2b^2}
	(R^\mu{}_{\nu}-{1\over 4}\delta^\mu_\nu R)  
	+{\chi^{\mu}{}_{\nu}(x)\over b^4}
	\label{sod1:trclss}  \ ,
\end{equation}
where the homogeneous solution satisfies the constraints 
\begin{eqnarray}
\chi^{\mu}{}_{\mu}=0   \ , \quad \chi^{\mu}{}_{\nu|\mu}=0 \ .
\label{sodTT}
\end{eqnarray}
 As we see later, this term  corresponds to  dark 
 radiation at this order.  
The metric can be obtained as
\begin{eqnarray}
  \overset{(1)}{g}{}_{\mu\nu} =  -{\ell^2 \over 2} 
  \left( {1\over b^2}-1 \right) 
    \left( R_{\mu\nu}  - {1\over 6} h_{\mu\nu} R \right)
      -{\ell \over 2}\left( {1\over b^4} -1 \right) \chi_{\mu \nu} \ ,
\end{eqnarray}
where we have imposed the boundary condition, 
$\overset{(1)}{g}{}_{\mu\nu} (y=0, x^\mu ) =0 $.

Let us focus on the role of  $\chi^\mu{}_{\nu}$ in this part.
 At this order, the junction condition can be written as
\begin{eqnarray}
\left[~\overset{(1)}{K}{}^{\mu}{}_{\nu} 
	-\delta^\mu_\nu\overset{(1)}{K}~\right] \Bigg|_{y=0}  
	&=& {\ell\over 2}\left(
	R^\mu{}_{\nu}-{1\over 2}\delta^\mu_\nu R
	\right)
	+\chi^\mu{}_{\nu}  \nonumber\\
	&=& {\kappa^2\over 2}T^\mu{}_{\nu}  \ .
\end{eqnarray}
Using the solutions Eqs.~(\ref{sod1:trc}), (\ref{sod1:trclss}) and the formula
\begin{equation}
E^\mu{}_{\nu}=K^\mu{}_{\nu ,y}-\delta^\mu_\nu K_{,y}
	-K^\mu{}_{\lambda}K^\lambda{}_{\nu}  
	+\delta^\mu_\nu K^\alpha{}_{\beta}K^\beta{}_{\alpha}
	-{3\over\ell^2}\delta^\mu_\nu   \ ,
\end{equation}
we  calculate the projective Weyl tensor as 
\begin{eqnarray}
\overset{(1)}{E}{}^{\mu}{}_{\nu}={2 \over\ell}\chi^\mu{}_{\nu} \ .
\end{eqnarray}
Then we obtain the effective Einstein equation
\begin{equation}
R^\mu{}_{\nu}-{1\over 2}\delta^\mu_\nu R  
	={\kappa^2\over\ell}T^\mu{}_{\nu}-\overset{(1)}{E}{}^{\mu}{}_{\nu}
	\label{sod1:effeq} \ .
\end{equation}
At this order, we do not have the conventional Einstein equations. 
 Recall that  the dark radiation exists even 
 in the low energy regime. Indeed, the low energy effective Friedmann
 equation becomes
\begin{equation}
  H^2 = {\kappa^2 \over 3\ell} \rho + {{\cal C} \over a^4}  \ .
\end{equation}
This equation can be obtained from Eq.~(\ref{sod1:effeq}) by
 imposing the maximal symmetry on the spatial part of the brane world
  and  the conditions (\ref{sodTT}).
Hence, we observe that $\chi^\mu{}_{\nu}$ is the generalization of 
the dark radiation found in the cosmological context. 

The nonlocal tensor $\chi_{\mu\nu}$ must be determined by the 
   boundary conditions in the bulk. 
 The natural choice is  asymptotically  AdS boundary condition.
 For this boundary condition, we have $\chi_{\mu\nu} =0 $. 
  It is this boundary condition that
  leads to the conventional Einstein theory in linearized gravity. 
Assuming this, we have
\begin{equation}
R^\mu{}_{\nu}-{1\over 2}\delta^\mu_\nu R  
	={\kappa^2\over\ell}T^\mu{}_{\nu}   \ .
\end{equation}
Thus, the conventional Einstein theory is recovered at the leading order!

\subsection{AdS/CFT Emerges}

In this subsection, we do not include the $\chi_{\mu\nu}$ field 
because we have adopted the AdS boundary condition. 
Of course, we have  calculated the second order solutions 
with the contribution of the $\chi_{\mu\nu}$ field.
 It merely adds  extra terms such as $\chi^\mu{}_{\nu}\chi^\nu{}_{\mu}$, etc. 
 
At  second order, the basic equations can be easily deduced. 
Substituting the solution up to  first order into the Ricci tensor
 and picking up the second order quantities, we obtain 
 the solutions at second order. 
The trace part is deduced algebraically as
\begin{eqnarray}
\overset{(2)}{K}
	= {\ell^3 \over 8 b^4} \left( R^\alpha{}_{\beta}R^\beta{}_{\alpha} 
	-{2\over 9}R^2 \right) 
	-{\ell^3 \over 12b^2} 
	\left( R^\alpha{}_{\beta} R^\beta{}_{\alpha} 
	-{1\over 6}R^2 \right)   \ .
\end{eqnarray}
 By integrating the equation for the traceless part, we have
\begin{eqnarray}
\overset{(2)}{\Sigma}{}^{\mu}{}_{\nu}  
	= -{\ell^2\over 2}\left( {y\over b^4} 
	+{\ell\over 2b^2} \right)
	{\cal S}^\mu{}_{\nu}  -{\ell^3 \over 24 b^2} 
	\left( RR^\mu{}_{\nu} 
	-{1\over 4}\delta^\mu_\nu R^2
	\right) 
	+{\ell^3 \over b^4}t^\mu{}_{\nu}   \ ,
\end{eqnarray}
where ${\cal S}^\mu{}_\nu$ is defined by 
\begin{eqnarray}
&&\delta \int d^4 x \sqrt{-h} {1\over 2} \left[ 
	R^{\alpha\beta}R_{\alpha\beta} -{1\over 3}R^2 \right] 
	=\int d^4 x \sqrt{-h}{\cal S}_{\mu\nu}
	\delta g^{\mu\nu} 
	\label{sodsmunu}\ .
\end{eqnarray}
The tensor ${\cal S}^\mu{}_\nu$ is transverse and traceless,
\begin{equation}
{\cal S}^\mu{}_{\nu |\mu} =0  \ ,\quad {\cal S}^\mu{}_{\mu} = 0   \ .
\end{equation}

The homogeneous solution $t^\mu{}_{\nu}$ must be traceless. 
 Moreover, it must satisfy the momentum constraint. 
 To be more precise, we must solve the constraint equation
\begin{equation}
t^\mu{}_{\nu |\mu}-{1\over 16}R^\alpha{}_{\beta}R^\beta{}_{\alpha|\nu}
	+{1\over 48}RR_{|\nu}-{1\over 24}R_{|\lambda}R^\lambda{}_{\nu}  
	=0 \ .
\end{equation}
As one can see immediately,
 there are ambiguities in integrating this equation. 
Indeed, there are two types of covariant local tensor whose
 divergences vanish:
\begin{eqnarray}
&&\delta\int d^4x\sqrt{-h}{1\over 2}R^{\alpha\beta}R_{\alpha\beta} 
	=\int d^4x\sqrt{-h}{\cal H}_{\mu\nu}\delta g^{\mu\nu}
	\label{sodhmunu} \ , \\
&&\delta\int d^4x\sqrt{-h}{1\over 2}R^2 
	=\int d^4x\sqrt{-h}{\cal K}_{\mu\nu}\delta g^{\mu\nu} 
	\label{sodkmunu}\ .
\end{eqnarray}
Notice that ${\cal S}^\mu{}_{\nu}={\cal H}^\mu{}_{\nu}
-{\cal K}^\mu{}_{\nu}/3 $. Hence, only ${\cal S}^\mu{}_\nu$ and
 ${\cal K}^\mu{}_\nu$ are independent. 
 Thanks to the Gauss-Bonnet topological 
 invariant, we do not need to consider the Riemann squared term.  
 In addition to these local tensors, we have to take into account
  the nonlocal tensor $\tau^\mu{}_{\nu}$ 
  with the property $\tau^\mu{}_{\nu|\mu}=0$. Thus, we get
\begin{eqnarray}
t^\mu{}_{\nu} 
	&=&{1\over 32}\delta^\mu_\nu 
	\left( R^\alpha{}_{\beta}R^\beta{}_{\alpha} 
	-{1\over 3}R^2 \right) 
	+{1\over 24}\left( RR^\mu{}_{\nu} 
         -{1\over 4}\delta^\mu_\nu R^2\right) \nonumber\\
	&&\qquad+\tau^\mu{}_{\nu}  
        +\left( \alpha+{1\over 4} \right){\cal S}^\mu{}_{\nu} 
        +{\beta\over 3}{\cal K}^\mu{}_{\nu}  \ , 
\end{eqnarray}
where the constants $\alpha$ and $\beta$ are free parameters
representing the degree of initial conditions in the bulk.
Hence, they represents the freedom of gravitational waves in the bulk. 
 The condition  $t^\mu{}_{\mu} =0$ leads to
\begin{equation}
\tau^\mu{}_{\mu}  
	=-{1\over 8}\left( R^\alpha{}_{\beta}R^\beta{}_{\alpha} 
        -{1\over 3}R^2 \right)-\beta\Box R      \ .
\end{equation}
This expression is the reminiscent of the trace anomaly of the CFT.
 It is possible to use the result of CFT at this point.  
 For example, we can choose the ${\cal N}=4$ super Yang-Mills theory
 as the conformal matter. In that case, we  simply put $\beta =0$. 
 This is the
 way the AdS/CFT correspondence comes into the brane world scenario. 

Up to the second order, the junction condition gives
\begin{eqnarray}
R^\mu{}_{\nu}-{1\over 2}\delta^\mu_\nu R 
	+2\ell^2\left[ \tau^\mu{}_{\nu} 
	+\alpha{\cal S}^\mu{}_{\nu}
	+{\beta\over 3}{\cal K}^\mu{}_{\nu}\right] 
	={\kappa^2\over\ell}T^\mu{}_{\nu}
	\label{sod2:effeq} \ .
\end{eqnarray}
 If we define 
\begin{eqnarray}
T_{\mu\nu}^{\rm CFT}= -2 {\ell^3 \over \kappa^2} \tau_{\mu\nu}  \ , 
\end{eqnarray}
we can write
\begin{eqnarray}
\overset{(4)}{G}{}_{\mu\nu} = {\kappa^2 \over \ell} \left( 
     T_{\mu\nu} + T_{\mu\nu}^{\rm CFT} \right) 
       -2 \ell^2  \alpha  {\cal S}_{\mu\nu} 
         - {2 \ell^2 \over 3} \beta {\cal K}_{\mu\nu}  \ .
\end{eqnarray}

Let us try to arrange the terms so as to reveal the geometrical
 meaning of the above equation. 
 We can calculate the projective Weyl  tensor as
\begin{equation}
\overset{(2)}{E}{}^{\mu}{}_{\nu} 
	=\ell^2\left[P^\mu{}_{\nu}+2\tau^\mu{}_{\nu} 
	+2\alpha{\cal S}^\mu{}_{\nu} 
	+{2\over 3}\beta{\cal K}^\mu{}_{\nu} \right]     \ ,
\end{equation}
where
\begin{eqnarray}
P^\mu{}_{\nu }&=&-{1\over 4}R^\mu{}_{\lambda}R^\lambda{}_{\nu}  
	+{1\over 6}R R^\mu{}_{\nu}
      +{1\over 8}\delta^\mu_\nu R^\alpha{}_{\beta}R^\beta{}_{\alpha}
	-{1\over 16}\delta^\mu_{\nu}R^2  \ .
\end{eqnarray}
Substituting this expression into Eq.~(\ref{sod2:effeq}) yields our main result
\begin{equation}
\overset{(4)}{G}{}_{\mu\nu}={\kappa^2\over\ell}T_{\mu\nu}   
	+\ell^2P_{\mu\nu}-\overset{(2)}{E}{}_{\mu\nu}
	\label{sod2:effeq2} \ .
\end{equation}
Notice that $E^\mu{}_{\nu}$ contains the nonlocal part and the free 
parameters $\alpha$ and $\beta$. 
 On the other hand, $P^\mu{}_{\nu}$ is determined locally. 
 One can see the relationship in a more transparent way.  
Within the accuracy we are considering,  
we can get $ P^\mu{}_{\nu }=\pi^\mu{}_{\nu}$ using the lowest order 
equation 
$R^\mu{}_{\nu}={\kappa^2 /\ell}(T^\mu{}_{\nu}-1/2\delta^\mu_\nu T) $.
 Hence, we can rewrite Eq.~(\ref{sod2:effeq2}) as 
\begin{equation}
\overset{(4)}{G}{}_{\mu\nu}={\kappa^2\over\ell}
	T_{\mu\nu}+\kappa^4\pi_{\mu\nu} 
	-\overset{(2)}{E}{}_{\mu\nu}
	\label{sod2:SMS} \ .
\end{equation}
 Now, the similarity between Eqs.~(\ref{sodShiMaSa}) and (\ref{sod2:SMS}) is 
 apparent. 
 Thus we get an explicit relation between the geometrical approach and the
 AdS/CFT approach.
However, we note that our Eq.~(\ref{sod2:SMS}) is a closed system
 of equations provided that the specific conformal field theory
 is chosen.

Now we can read off the effective action as
\begin{eqnarray}
S_{\rm eff}&=&\frac{\ell}{2\kappa^2}\int d^4x\sqrt{-h}~R
	+S_{\rm matter}+S_{\rm CFT}  \nonumber \\
	&&\quad
	+\frac{\alpha\ell^2}{2\kappa^2}
	\int d^4x\sqrt{-h}
	\left[R^{\mu\nu}R_{\mu\nu}
	-{1\over 3}R^2 \right] \nonumber \\
	&&\quad
	+\frac{\beta\ell^2}{6\kappa^2}\int d^4x\sqrt{-h}~R^2  \ ,
\end{eqnarray}
where we have used the relations Eqs.~(\ref{sodsmunu}), (\ref{sodhmunu}) 
and (\ref{sodkmunu}) and we denoted the 
nonlocal effective action constructed from $\tau^\mu{}_{\nu}$ as $S_{\rm CFT}$.

\section{Two-brane model (RS1)}

In this section, we will apply the gradient expansion mrthod to the
two-brane models and reveal a role of the AdS/CFT correspondence. 

\begin{figure}[h]
\centerline{\includegraphics[width = 6cm, height = 7cm]{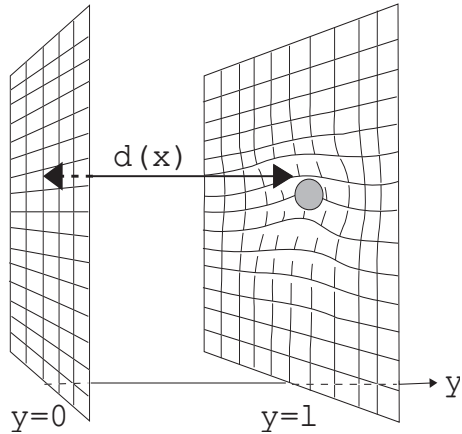}}
\caption{In the two brane system, the radion is defined
as a distance between two branes. It could depend on the position because
the brane could have bending. Hence, 
for observers on the brane, it appears as a scalar field. }
\end{figure}

\subsection{Scalar-Tensor Theory Emerges}

 We consider  the two-brane system depicted in Fig.2.
 Without matter on the branes, we have the relation
$g^{\ominus\hbox{-}\rm brane}_{\mu\nu}
=e^{-2d/\ell}g^{\oplus\hbox{-}\rm brane}
\equiv \Omega^2 g^{\oplus\hbox{-}\rm brane} $
where $d$ is the distance between the two branes.
Although $\Omega$ is constant for vacuum branes, it becomes the
function of the 4-dimensional coordinates if we put the matter on the brane.

Adding the energy momentum tensor to each of the two branes,
and allowing deviations from the pure AdS$_5$ bulk, the
effective (non-local) Einstein equations on the branes at low
energies take the form\cite{sodKS2},
\begin{eqnarray}
G^\mu{}_{\nu} (h )
	&=&{\kappa^2\over\ell}\overset{\oplus}{T}{}^{\mu}{}_{\nu} 
	-{2\over \ell}\chi^\mu{}_{\nu} \,,
\label{sodA:einstein}
\\
G^\mu{}_{\nu}(f)
&=&-{\kappa^2\over\ell}\overset{\ominus}{T}{}^{\mu}{}_{\nu} 
	-{2\over\ell}{\chi^\mu{}_{\nu}\over\Omega^4} \ .
\label{sodB:einstein}
\end{eqnarray}
where $h_{\mu\nu}=g^{\oplus\hbox{-}{\rm brane}}_{\mu\nu}$, 
$f_{\mu\nu}=g^{\ominus\hbox{-}{\rm brane}}_{\mu\nu}=\Omega^2h_{\mu\nu}$
and the terms proportional to $\chi_{\mu\nu}$ are 5-dimensional
Weyl tensor contributions which describe
the non-local 5-dimensional effect.
Although Eqs.~(\ref{sodA:einstein}) and (\ref{sodB:einstein})
are non-local individually, with undetermined $\chi_{\mu\nu}$,
one can combine both equations to reduce them to local equations
for each brane. Since $\chi_{\mu\nu}$ appears only algebraically,
one can easily eliminate $\chi_{\mu\nu}$ from Eqs.~(\ref{sodA:einstein}) 
and (\ref{sodB:einstein}). 
Defining a new field $\Psi = 1-\Omega^2$,  we find 
\begin{eqnarray}
\hspace{-3mm}
G^\mu{}_{\nu}(h)&=&{\kappa^2 \over \ell \Psi } 
	\overset{\oplus}{T}{}^{\mu}{}_{\nu}
	+{\kappa^2 (1-\Psi )^2 \over \ell\Psi } 
	\overset{\ominus}{T}{}^{\mu}{}_{\nu} \nonumber \\
&&	+{1\over\Psi}\left(\Psi^{|\mu}{}_{|\nu} 
  	-\delta^\mu_\nu\Psi^{|\alpha}{}_{|\alpha}\right) \nonumber\\
&&	+{3 \over 2 \Psi (1-\Psi )}\left( \Psi^{|\mu}\Psi_{|\nu}
  	- {1\over 2}\delta^\mu_\nu\Psi^{|\alpha}\Psi_{|\alpha} 
  	\right),
  	\label{sodA:STG1} \\
\hspace{-3mm}
\Box\Psi&=&{\kappa^2\over 3\ell}(1-\Psi )
	\left\{ \overset{\oplus}{T} + (1-\Psi)\overset{\ominus}{T}  
	                     \right\} 
       	-{1 \over 2 (1-\Psi )} \Psi^{|\mu}\Psi_{|\mu} \ , 
  	\label{sodA:STG2}  	
\end{eqnarray}
where $|$ denotes the covariant derivative with respect to the metric
$h_{\mu\nu}$. Since $\Omega$ (or equivalently $\Psi$) contains
the information of the distance between the two branes,
we call $\Omega$ (or $\Psi$) the radion.

We can also determine $\chi^{\mu}{}_{\nu}$ by eliminating $G^{\mu}{}_{\nu}$ 
from Eqs.~(\ref{sodA:einstein}) and (\ref{sodB:einstein}). Then,
\begin{eqnarray}
\chi^{\mu}{}_{\nu}&=&-{\kappa^2(1-\Psi)\over 2 \Psi} 
	\left( \overset{\oplus}{T}{}^{\mu}{}_{\nu} 
	+(1-\Psi)\overset{\ominus}{T}{}^{\mu}{}_{\nu}\right)  \nonumber\\
&&	-{\ell\over 2\Psi} \left[ \left(  \Psi^{|\mu}{}_{|\nu} 
	-\delta^\mu_\nu  \Psi^{|\alpha}{}_{|\alpha} \right) 
	\right. \nonumber \\
&&	\left.+{3 \over 2(1 -\Psi )} \left( \Psi^{|\mu}  \Psi_{|\nu}
  	-{1\over 2} \delta^\mu_\nu  \Psi^{|\alpha} \Psi_{|\alpha} 
  	\right) \right]   \ .  
  	\label{sodA:chi}
\end{eqnarray}
 Note that the index of $\overset{\ominus}{T}{}^{\mu}{}_{\nu}$ is to be raised
or lowered by the induced metric on the $\ominus$-brane, $f_{\mu\nu}$.

The effective action for the $\oplus$-brane which gives 
Eqs.~(\ref{sodA:STG1}) and (\ref{sodA:STG2}) is
\begin{eqnarray}
S_{\oplus}&=&{\ell \over 2 \kappa^2} \int d^4 x \sqrt{-h} 
	\left[ \Psi R - {3 \over 2(1- \Psi )} 
     	\Psi^{|\alpha} \Psi_{|\alpha} \right] \nonumber\\
    && \!\!\!\!\!\!\!\!\!\!
    	+ \int d^4 x \sqrt{-h} {\cal L}^\oplus 
      	+ \int d^4 x \sqrt{-h} \left(1-\Psi \right)^2 {\cal L}^\ominus  
      	\ .  
      	\label{sodA:action} 
\end{eqnarray}
The above action can be used to make cosmological predictions~\cite{sodKoyama:2003be}.
It should be stressed that  the radion has the conformal coupling. 
In fact, using the transformation $\Psi = 1- \psi^2$, we obtain
\begin{eqnarray}
 S_{\oplus}={6 \ell \over  \kappa^2} \int d^4 x \sqrt{-h} 
	\left[ \frac{1}{12} R -\frac{1}{12} \psi^2 R - \frac{1}{2} 
     	\psi^{|\alpha} \psi_{|\alpha} \right] + \cdots
      	\ .  
\end{eqnarray}
This is nothing but Einstein theory with 
a conformally coupled scalar field $\psi$.
 
\subsection{AdS/CFT in two-brane system?}

 In the two-brane case, it is difficult to proceed to the next order 
 calculations. Hence, we need to invent a new method~\cite{sodKanno:2003vf}. 
 For this purpose, we shall start with the effective Einstein equation obtained 
 by Shromizu, Maeda, and Sasaki~\cite{sodShiMaSa}
\begin{eqnarray}
 G_{\mu\nu} = T_{\mu\nu} + \pi_{\mu\nu} - E_{\mu\nu}
\end{eqnarray}
where $\pi_{\mu\nu}$ is the quadratic of energy momentum tensor $T_{\mu\nu}$ and
$E_{\mu\nu}$ represents the  effect of the bulk geometry. 
 Here we have set $8\pi G=1$. 
 This geometrical projection approach can not give a concrete prediction, 
 because we do not know $E_{\mu\nu}$ without solving the equations of motion 
 in the bulk. Fortunately, in the case of the homogeneous cosmology, 
 the property $E^\mu{}_{\mu} =0$ determines the dynamics as 
\begin{eqnarray}
  H^2 = {1  \over 3} \rho + \rho^2 + {{\cal C} \over a^4} \ .
\end{eqnarray}
 This reflects
 the interplay between the bulk and the brane dynamics on the brane.  

 What we want to seek is an effective theory which contains the information
 of the bulk as finite number of constant parameters like ${\cal C}$ in the
 homogeneous universe. When we succeed in obtaining it, the cosmological 
 perturbation theory can be constructed in a usual way. Although the
 concrete prediction can not be made, qualitative understanding of
 the evolution of the cosmological fluctuations can be obtained.
 This must be useful to make observational predictions.

 In the two-brane system, the mass spectrum is known from the linear
 analysis~\cite{sodGT}. At low energy, the propagator for the KK mode
 with the mass $m$ can be expanded as 
\begin{eqnarray}
   {-1 \over \Box - m^2} = {1\over m^2}\left[ 1 + {\Box \over m^2}
   + {\Box^2 \over m^4}+ \cdots \right] \ .
\end{eqnarray}
 However, massless modes can not be expanded in this way, hence we must 
 take into account all of the massless modes to construct braneworld
 effective action. 
 It seems legitimate to assume this consideration is valid even 
  in the non-linear regime.  Thus, at low energy,
 the action can be expanded by the local terms with increasing
  orders of derivatives of the metric $g_{\mu\nu}$ 
  and the radion $\Psi$~\cite{sodKS2}.

 Let us illustrate our method using the following action truncated at the second
 order derivatives:
\begin{eqnarray}
 S_{\rm eff} 
 	={1\over 2}  \int d^4 x \sqrt{-g} \left[ 
 	\Psi R -2\Lambda (\Psi)   
	-{\omega (\Psi) \over \Psi} 
	\nabla^\mu\Psi\nabla_\mu\Psi\right]
	\label{sod1:action} \ ,
\end{eqnarray}
 which is nothing but the scalar-tensor theory with coupling function
 $\omega (\Psi)$ and the potential function $\Lambda (\Psi )$. 
 Note that this is the most general local action which contains 
 up to the second  order derivatives and has the general coordinate invariance.
 It should be stressed that the scalar-tensor theory is, in general,
  not related to the braneworld. However, we know a special type of 
 scalar-tensor theory corresponds to  the low energy 
 braneworld~\cite{sodKS2, sodShiKo, sodSK,sodwiseman,sodchiba,sodKobayashi:2002pw}. 
 Here, we will present a simple derivation of this known fact.  
 
 For the vacuum brane, we can put 
$T_{\mu\nu} + \pi_{\mu\nu} = - \lambda g_{\mu\nu}$. Hence, 
 the geometrical effective equation  reduces to 
\begin{eqnarray}
G_{\mu\nu}=-E_{\mu\nu}-\lambda g_{\mu\nu}
\label{sod1:SMS}\ .
\end{eqnarray}
 First, we must find $E_{\mu\nu}$. 
The above action (\ref{sod1:action}) gives the equations of motion for 
the metric as
\begin{eqnarray}
G_{\mu\nu}&=&-{\Lambda \over \Psi} g_{\mu\nu}
	+{1\over \Psi} \left( 
	\nabla_\mu \nabla_\nu \Psi
	-g_{\mu\nu} \Box \Psi \right)
	\nonumber\\
&&	+ {\omega \over \Psi^2} \left(
        \nabla_\mu \Psi \nabla_\nu \Psi -{1\over 2}g_{\mu\nu}
        \nabla^\alpha \Psi \nabla_\alpha\Psi\right)
        \label{sod1:ST}  \ . 
\end{eqnarray}
The right hand side of this Eq.~(\ref{sod1:ST}) should be identified with 
$-E_{\mu\nu}-\lambda g_{\mu\nu}$.
 Hence, the  condition  $E^\mu{}_\mu =0$ becomes
\begin{eqnarray}
\Box\Psi = - {\omega \over 3\Psi} 
        \nabla^\mu \Psi \nabla_\mu \Psi  
        - {4\over 3} \left( \Lambda  - \lambda \Psi \right)  \ .
	\label{sod1:KG1}
\end{eqnarray}
This is the equation for the radion $\Psi$. However, we also
have the equation for $\Psi$ from the action (\ref{sod1:action}) as 
\begin{eqnarray}
\Box \Psi = \left( {1\over 2\Psi} - { \omega' \over 2\omega} \right)
	\nabla^\alpha \Psi \nabla_\alpha \Psi  
	-{\Psi \over 2\omega} R + {\Psi \over \omega} \Lambda'    \ ,
	\label{sod1:KG2}
\end{eqnarray}
where the prime denotes the derivative with respect to $\Psi$. 
In order for these two Eqs.~(\ref{sod1:KG1}) and (\ref{sod1:KG2}) to be compatible, 
$\Lambda$ and $\omega$ must satisfy 
\begin{eqnarray}
&&-{\omega\over 3\Psi}
	={1\over 2\Psi}-{ \omega' \over 2\omega}
	\label{sodeq1}
	\ ,  \\
&&{4\over 3} \left( \Lambda-\lambda\Psi\right)
	={\Psi\over\omega}\left( 2\lambda -\Lambda'\right)
	\label{sodeq2}  \ ,
\end{eqnarray}
where we used  $R= 4\lambda$ which comes from the trace part of 
Eq.~(\ref{sod1:SMS}). Eqs.~(\ref{sodeq1}) and (\ref{sodeq2}) can be integrated as  
\begin{eqnarray}
   \Lambda (\Psi) = \lambda + \lambda \gamma \left( 1-\Psi \right)^2   \ , \quad
  \omega (\Psi ) = {3\over 2} {\Psi \over 1-\Psi}  \ ,  
\end{eqnarray}
where the constant of integration $\gamma$ represents the ratio
 of the cosmological constant on the negative tension brane to that on 
 the positive tension brane. 
 Here, one of constants of integration is absorbed by rescaling of $\Psi$.
 In doing so, we have assumed the constant of integration is positive.
 We can also describe the negative tension brane if we take the 
 negative signature.

Thus, we get the effective action 
\begin{eqnarray}
S_{\rm eff}
	=\int d^4 x \sqrt{-g} \left[ {1\over 2} \Psi R 
	-{3 \over 4( 1-\Psi )} \nabla^\mu \Psi \nabla_\mu \Psi 
	- \lambda - \lambda \gamma (1-\Psi)^2 \right] \ .
\end{eqnarray}
Surprisingly, this completely agrees with the previous 
result (\ref{sodA:action}). Our simple symmetry 
principle $E^\mu{}_\mu =0$ has determined the action completely. 
 
 As we have shown in \cite{sodKSS}, if $\gamma <-1$
 there exists a static deSitter two-brane solution
 which turns out to be unstable. In particular,  
 two inflating branes  can collide at $\Psi = 0$. 
 This process is completely smooth for the observer on the brane.  
 This fact led us to the born-again scenario~\cite{sodKSS,sodKanno:2005vq}.
 The similar process occurs also in the ekpyrotic (cyclic) model~\cite{sodturok}
 where the moduli approximation is used. It can be shown that the moduli
 approximation is nothing but the lowest order truncation of the low energy
 gradient expansion method developed 
 by us~\cite{sodKanno:2004yb,sodKanno:2005zr,sodBrax:2002nt,sodMcFadden:2004ni}. 
 Hence, it is of great interest to see the leading order corrections due
 to KK modes to this process.  

  Let us now apply the conformal symmetry method explained above 
 to the higher order case. 
 First, we need to write down the most generic action containing up to fourth order 
 derivatives. Then, we impose the symmetry to determine unknown functionals.
 The action reads
\begin{eqnarray}
S_{\rm eff}
\!\!&=&\!\!
	{1\over 2} \int d^4 x \sqrt{-g} \left[ \Psi R - 2\Lambda (\Psi )
	-{\omega (\Psi) \over \Psi} \nabla^\mu \Psi \nabla_\mu \Psi \right]
        \nonumber\\
&&\!\!
	+\int d^4 x \sqrt{-g} \left[
	A(\Psi) \left( \nabla^\mu \Psi \nabla_\mu \Psi \right)^2
	+B(\Psi) \left( \Box \Psi \right)^2  \right. \nonumber\\
&&\left. \quad\qquad	+C(\Psi)\nabla^\mu \Psi \nabla_\mu \Psi \Box \Psi
	+D(\Psi) R~\Box \Psi 
	      \right. \nonumber\\
&&\left. \quad\qquad
	+ E(\Psi) R \nabla^\mu \Psi \nabla_\mu \Psi
        + F(\Psi) R^{\mu\nu} \nabla_\mu \Psi \nabla_\nu \Psi  
              \right. \nonumber\\
&&\left. \quad\qquad
        + G(\Psi) R^2     
	+ H(\Psi) R^{\mu\nu} R_{\mu\nu} 
	        \right. \nonumber\\
&&\left. \quad\qquad
	+I(\Psi) R^{\mu\nu\lambda\rho} R_{\mu\nu\lambda\rho} 
	+\cdots  \right]  \ ,
	\label{sodsetup}
\end{eqnarray}
where $A,B, \cdots$ denote arbitrary functionals of the radion.  

   Now we impose the conformal symmetry  on the fourth order derivative terms
  in the action (\ref{sodsetup}) as we did in the  previous example. 
  Starting from the action (\ref{sodsetup}), one can read off the equation for 
  the metric from which $E_{\mu\nu}$ can be identified. 
  The compatibility between the equations of motion for $\Psi$
  and the equation $E^\mu{}_\mu =0$ constrains the coefficient functionals
  in the action (\ref{sodsetup}). Surprisingly, every coefficient functionals are
  determined up to constants. 
 
Thus, we find the 4-dimensional effective action with KK corrections as
\begin{eqnarray}
S_{\rm eff} 
	&=& \int d^4 x \sqrt{-g} \left[ {1\over 2} \Psi R 
	-{3 \over 4( 1-\Psi )} \nabla^\mu \Psi \nabla_\mu \Psi 
        \right. \nonumber\\
&&\left.\quad
	-\lambda-\lambda\gamma (1-\Psi)^2
	\right]\nonumber\\
&&	+\ell^2 \int d^4 x \sqrt{-g}
	\left[
	{1 \over 4 (1-\Psi)^4} \left( \nabla^\mu \Psi \nabla_\mu \Psi \right)^2
	\right. \nonumber\\
&&\left.
	+{1\over (1-\Psi)^2} \left( \Box \Psi \right)^2 
	+{1\over (1-\Psi)^3} \nabla^\mu \Psi \nabla_\mu \Psi\Box \Psi
	\right. \nonumber \\
&&\left.
	+ {2\over 3(1-\Psi)} R \Box \Psi 
	+{1\over 3(1-\Psi)^2}  R \nabla^\mu \Psi \nabla_\mu \Psi 
	\right. \nonumber\\
&&\left.
	+jR^2+ k R^{\mu\nu} R_{\mu\nu}  \right] \ ,
	\label{sodaction}
\end{eqnarray}
where constants $j$ and $k$ can be interpreted as 
 the variety of the effects of the bulk gravitational waves.
 These constants have the same origin as the previous
 parameters $\alpha$ and $\beta$.
It should be noted that this action becomes non-local after
 integrating out the radion field. This fits the fact that
 KK effects\index{KK effects} are non-local usually.  
  In principle, we can continue this calculation to any order of 
 derivatives.

\section{The Answers}

We have developed the low energy gradient expansion scheme
 to give  insights into the physics of the braneworld
 such as the black hole physics and the cosmology.
 In particular, we have concentrated on the specific questions 
 in this lecture.
 Here, we summarize our answers obtained by the gradient expansion method.
 Our understanding of RS braneworlds would be also useful for other brane models.

\subsection{Single-brane model}

\noindent
{\bf Is the Einstein theory  recovered even in the non-linear regime?}\\

 We have obtained the effective theory at the lowest order
as
\begin{equation}
\overset{(4)}{G}{}^\mu{}_{\nu}  
	={\kappa^2\over\ell}T^\mu{}_{\nu} 
	-{2\over \ell} \chi^{\mu}{}_{\nu} \ .
\end{equation}
Here we have the correction $\chi_{\mu\nu}$ 
which can be interpreted as the dark radiation 
in the cosmological situation. 

 On the other hand, in the linearized gravity, 
 the conventional Einstein theory is recovered at
 low energy. This is because the out-going boundary condition is imposed.
 In other words, the asymptotic AdS boundary condition is imposed.
 In the nonlinear case, this corresponds to the requirement that
  the dark radiation term $\chi_{\mu\nu}$ must be zero. 
 For this boundary condition, the conventional Einstein theory is recovered.
 Hence, the standard Friedmann equation holds. 
 
 In this sense, the answer is yes.

\vskip 0.5cm
\noindent
{\bf How does the AdS/CFT come into the braneworld?}\\

The CFT emerges as the constant of integration which satisfies
the trace anomaly relation
\begin{equation}
\tau^\mu{}_{\mu}  
	=-{1\over 8}\left( R^\alpha{}_{\beta}R^\beta{}_{\alpha} 
	-{1\over 3}R^2\right)-\beta\Box R      \ .
\end{equation}
This constant can not be determined a priori. 
 Here, the AdS/CFT correspondence could come into the braneworld.
 Namely, if we identify some CFT with $\tau_{\mu\nu}$, then we can
 determine the boundary condition.

\vskip 0.5cm
\noindent
{\bf How are the AdS/CFT and geometrical approach  related?}\\

The key quantity in the geometric approach is obtained as
\begin{equation}
\overset{(2)}{E}{}^{\mu}{}_{\nu} 
	=\ell^2\left[P^\mu{}_{\nu}+2\tau^\mu{}_{\nu} 
	+2\alpha{\cal S}^\mu{}_{\nu} 
	+{2\over 3}\beta{\cal K}^\mu{}_{\nu}\right]     \ .
\end{equation}
The above expression contains $\tau_{\mu\nu}$ which can be interpreted
 as the CFT matter. Hence, once we know $E_{\mu\nu}$, no enigma remains.
 In particular, $P_{\mu\nu}\approx \pi_{\mu\nu}$ is independent
 of the $\tau_{\mu\nu}$. In odd dimensions, there exists
 no trace anomaly, but $P_{\mu\nu}$ exists. In 4-dimensions, 
 $\pi^\mu{}_\mu$ accidentally coincides with the trace anomaly in CFT. 
 
 It is interesting to note that the high energy and 
 the Weyl term corrections found in the geometrical
    approach merge into the CFT matter correction found in the 
    AdS/CFT approach.

\subsection{Two-brane model}

\noindent
{\bf How is the geometrical approach consistent with the Brans-Dicke
 picture?}\\

 In the geometrical approach, no radion seems to appear. On the other
 hand, the linear theory predicts the radion as the crucial quantity.
 The resolution can be attained by obtaining $E_{\mu\nu}$
 ($\chi_{\mu\nu}$ in our notation). The resultant expression
\begin{eqnarray}
\chi^{\mu}{}_{\nu}&=&-{\kappa^2(1-\Psi)\over 2 \Psi} 
	\left( \overset{\oplus}{T}{}^{\mu}{}_{\nu} 
	+(1-\Psi)\overset{\ominus}{T}{}^{\mu}{}_{\nu}\right)  \nonumber\\
&&	-{\ell\over 2\Psi} \left[ \left(  \Psi^{|\mu}{}_{|\nu} 
	-\delta^\mu_\nu  \Psi^{|\alpha}{}_{|\alpha} \right) 
	\right. \nonumber \\
&&	\left.+{3 \over 2(1 -\Psi )} \left( \Psi^{|\mu}  \Psi_{|\nu}
  	-{1\over 2} \delta^\mu_\nu  \Psi^{|\alpha} \Psi_{|\alpha} 
  	\right) \right]   \nonumber
\end{eqnarray}
contains the radion in an intriguing way. 
The dark radiation consists of the radion and the matter. 

 We have shown that the radion transforms the Einstein theory with
  Weyl correction into the conformally coupled scalar-tensor theory 
  where the radion plays the role of the scalar field.
 Thus, it turned out that the radion is hidden by 
 the projected Weyl tensor $E_{\mu\nu}$ in the geometrical approach.

\vskip 0.5cm
\noindent
{\bf What replaces  the AdS/CFT correspondence 
in the two-brane model?}\\

 In the case of the single-brane model, the out-going boundary condition 
 at the Cauchy horizon is assumed. This conforms to AdS/CFT correspondence.
 Indeed, the continuum KK-spectrum are projected on the brane as CFT matter.
\begin{figure}[h]
\centerline{\includegraphics[width = 6cm, height = 7cm]{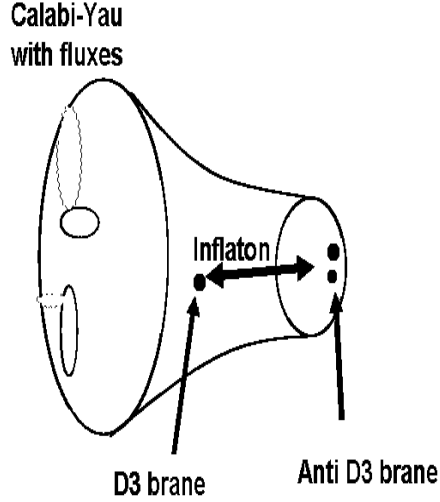}}
\caption{The warped throat is attached to Calabi-Yau manifold.
Anti-D3-branes are stacked at the tip of the warped throat.
A D3-brane can move in the throat and the radion, i.e. the
 distance between the D3-brane and anti-D3-branes, plays a role of the inflaton.}
\end{figure}

 On the other hand, the boundary condition in the two-brane system 
 allows only the discrete KK-spectrum. Hence, we can not expect CFT
 matter on the brane. Instead, the radion controls the bulk/brane
 correspondence in two-brane model. 
 In fact, the higher derivative terms of the radion mimics
 the effect of the bulk geometry (KK-effect) as we have shown explicitly. 
 Hence, the conventional AdS/CFT correspondence does not exist. 
 Instead, there exists the AdS/CFT correspondence realized by the conformally
 coupled radion. The conformal coupling can be regarded as
  a reflection of the symmetry of the bulk geometry.

Here, I would like to mention D-brane inflation\index{D-brane inflation} models
proposed in \cite{sodKachru:2003sx}.
There, the inflaton is identified as the radion which is the distance
between a D3-brane and anti-D3-branes as is shown in Fig.3.
 Naively, it seems possible to
realize a slow roll inflation due to the warped geometry.
However,  our result of two-brane system suggests 
the existence of the conformal coupling of the radion,
which ruins the slow role inflation of the model. In fact, the curvature coupling
gives a large mass of the inflaton which causes the notorious eta problem. 
Hence, a fine tuning is unavoidable.

\section{AdS/CFT in Dilatonic braneworld}

 In the previous sections, we have considered RS braneworlds.
 If we take into account bulk fields, the pure AdS bulk would not be
 expected. Hence, it is interesting to see the role of AdS/CFT
 correspondence in those cases. In this section, we will consider a bulk
 scalar field and call this kind of models dilatonic braneworlds~\cite{sodMaeda:2000wr}.
 
 In addition to the above theoretical interest, there is an phenomenological
 intererest in dilatonic braneworlds. 
 Let us see how the inflationary universe
 can be realized in the braneworld. 
 The formula for the effective cosmological constant
in the braneworld reads 
\begin{eqnarray}
    \Lambda_{\rm eff} = {\kappa^4 \sigma^2  \over 12}
                      - {3 \over \ell^2 } \ ,
\end{eqnarray}
where $\sigma$ and $\ell$ are the tension of the brane and the curvature
scale in the bulk which is determined by the bulk vacuum energy, respectively.
 For $\kappa^2 \sigma = 6/\ell$, we have Minkowski spacetime.
 In order to obtain the inflationary universe, we need the positive
 effective cosmological constant. In the braneworld model, there are
 two possibilities. One is to increase the brane tension and the other
 is to increase $\ell$. The brane tension can be controlled by the scalar
 field on the brane. The bulk curvature scale $\ell$ can be controlled
  by the bulk scalar field. The former case is a natural extension of the
  4-dimensional inflationary scenario. The latter possibility
  is a novel one peculiar to the brane model. 
   Recall that, in the superstring theory,  scalar fields  are 
 ubiquitous. Indeed, the dilaton and moduli exists in the bulk 
 generically, because they arise as the  modes associated with the 
 closed string. Moreover, when the supersymmetry is spontaneously broken,
 they may have the non-trivial potential. Hence, it is natural to consider 
 the inflationary scenario driven by these 
 fields~\cite{sodKobayashi:2000yh,sodHimemoto:2000nd}. 
 Therefore, dilatonic braneworlds are phenomenologically interesting.
 In this section, we would like to discuss this dilatonic braneworld\index{dilatonic braneworld}
  from the point of view of the AdS/CFT correspondence.

\subsection{Dilatonic Braneworld}
 
 We consider a $S_1/Z_2$ orbifold spacetime with the two branes 
as the fixed points. In this first Randall-Sundrum (RS1) model, 
the two flat 3-branes are embedded in AdS$_5$ 
 and the brane tensions given by
$\overset{\oplus}{\sigma}=6/(\kappa^2\ell)$ and 
$\overset{\ominus}{\sigma}=-6/(\kappa^2\ell)$. 
 Our system is described by the action 
\begin{eqnarray}
S &=& {1\over 2\kappa^2}\int d^5 x \sqrt{-g}~{\cal R}
	-\int d^5 x  \sqrt{-g} \left[~ 
	{1\over 2} g^{AB} \partial_A\varphi 
	\partial_B\varphi
	+ U(\varphi) ~\right]\nonumber\\
&&	-\sum_{i=\oplus,\ominus} \overset{i}{\sigma} 
\int d^4 x \sqrt{-g^{i\mathrm{\hbox{-}brane}}} 
+\sum_{i=\oplus,\ominus} \int d^4 x \sqrt{-g^{i\mathrm{\hbox{-}brane}}}
\,{\cal L}_{\rm matter}^i  \ ,
\label{sod5D:action}
\end{eqnarray}
where  $g^{i\mathrm{\hbox{-}brane}}_{\mu\nu}$ 
and $\overset{i}{\sigma}$ are  the induced metric 
 and the brane tension on the $i$-brane, respectively. 
We assume the potential $U(\varphi)$ for the bulk scalar field takes the form
$
U(\varphi)=-\frac{6}{\kappa^2\ell^2}+V(\varphi)\ ,
\label{sodpotential}
$
where the first term is regarded as a 5-dimensional cosmological constant
and the second term is an arbitrary potential function.
The brane tension $\sigma$ is tuned so that
the effective cosmological constant on the brane vanishes. 
The above setup realizes a flat braneworld after inflation ends and 
the field $\varphi$ reaches the minimum of its potential. 

Inflation in the braneworld can be driven by a scalar field 
either on the brane or in the bulk.
We derive the effective equations of motion which are useful for both 
models. 
Here, we begin with the single-brane system. Since we 
know the effective 4-dimensional equations hold irrespective of the existence
 of other branes~\cite{sodKS4},
  the analysis of the single-brane system is sufficient
 to derive the effective action for the two-brane system 
 as we see in the next subsection.

We again adopt the Gaussian normal coordinate system to describe 
the geometry of the brane model;
$
ds^2 = dy^2 + g_{\mu\nu} (y,x^{\mu} ) dx^{\mu} dx^{\nu}  \ ,
$
where the brane is assumed to be located at $y=0$. 
 Let us decompose the extrinsic curvature into the traceless part 
 $\Sigma_{\mu\nu}$ and the trace part $K$ as
$
K_{\mu\nu}=-\frac{1}{2}g_{\mu\nu,y}
	=\Sigma_{\mu\nu}
	+{1\over 4} g_{\mu\nu} K \  .
$
Then, we can obtain the basic equations off the brane 
using these variables.
First, the Hamiltonian constraint equation leads to
\begin{eqnarray}       		
{3\over 4} K^2-\Sigma^{\alpha}{}_{\beta}\Sigma^{\beta}{}_{\alpha}
	&=& \overset{(4)}{R} -\kappa^2 \nabla^\alpha\varphi
	\nabla_\alpha\varphi 
	+\kappa^2(\partial_y\varphi )^2
	-2\kappa^2U(\varphi ) \ ,  
    	\label{sodeq:hamiltonian} 
\end{eqnarray}
where $\overset{(4)}{R}$ is the curvature on the brane and 
$\nabla_\mu $ denotes the
covariant derivative with respect to the metric $g_{\mu\nu}$. Momentum 
constraint equation becomes
\begin{eqnarray}       		
\nabla_\lambda \Sigma_{\mu}{}^{\lambda}  
	-{3\over 4} \nabla_\mu K =-\kappa^2
	\partial_y\varphi\partial_\mu\varphi\ .
	\label{sodeq:momentum}  
\end{eqnarray}
Evolution equation in the direction of $y$ is given by
\begin{eqnarray}
\Sigma^{\mu}{}_{\nu,y}-K\Sigma^{\mu}{}_{\nu} 
	=-\left[ \overset{(4)}{R}{}_{\mu\nu}-\kappa^2 \nabla^\mu\varphi
	\nabla_\nu\varphi 
	\right]_{\rm traceless}      \ .     
	\label{sodeq:evolution} 
\end{eqnarray}
Finally,  the equation of motion for the scalar field reads
\begin{eqnarray}
\partial^2_y \varphi -K\partial_y\varphi 
	+ \nabla^{\alpha}\nabla_{\alpha}\varphi -U'(\varphi)=0  \ ,
	\label{sodeq:scalar}	
\end{eqnarray}
where the prime denotes derivative with respect to the scalar field $\varphi$.

 As we have the singular source at the brane position, we must consider 
 the junction conditions. 
 Assuming a $Z_2$ symmetry of spacetime, we obtain the junction 
conditions for the metric and the scalar field 
\begin{eqnarray}
\left[ \Sigma^{\mu}{}_{\nu}-{3\over 4} \delta^\mu_\nu K \right] \Bigg|_{y=0}
	&=& -{\kappa^2 \over 2}\sigma  \delta^\mu_\nu 
    	+\frac{\kappa^2}{2}T^{\mu}{}_{\nu}  \ , 
    	\label{sodJC:metric} \\
	\Bigl[\partial_y\varphi\Bigr]\bigg|_{y=0} &=& 0 \ ,
	\label{sodJC:scalar}         
\end{eqnarray}
where $T^{\mu}{}_{\nu}$ is the energy-momentum tensor for the matter
fields  on the brane.

\subsection{AdS/Radion correspondence}

 We assume the inflation occurs at low energy
 in the sense that the additional energy due to the bulk scalar field is small, 
 $\kappa^2\ell^2V(\varphi)\ll 1$, and the curvature on the brane $R$ 
 is also small, $R\ell^2\ll 1$.  It should be stressed that the low energy
 does not necessarily implies weak gravity on the brane. 
 Under these circumstances, we can use a
 gradient expansion scheme to solve the bulk equations of motion.  

 At zeroth order,  we ignore matters on the brane.
  Then, from the junction condition (\ref{sodJC:metric}), we have 
\begin{eqnarray} 
\left[ \overset{(0)}{\Sigma}{}^{\mu}{}_{\nu}
 -{3\over 4} \delta^\mu_\nu \overset{(0)}{K} \right] \Bigg|_{y=0}
	&=& -{\kappa^2 \over 2}\sigma \delta^\mu_\nu  \ .
	\label{sodJC:metric0}
\end{eqnarray}
 As the right hand side of (\ref{sodJC:metric0}) contains no traceless part,
 we get 
$
\overset{(0)}{\Sigma}{}^\mu{}_\nu  =0 \ .
$
We also take the potential for the bulk scalar field 
$U(\varphi)$ to be $-6/(\kappa^2\ell^2)$. 
We discard the terms with  4-dimensional derivatives since one can neglect
 the long wavelength variation in the direction of $x^\mu$ at low energies.  
 Thus, the equations to be solved are given by
\begin{eqnarray}       		
  && {3\over 4} \overset{(0)}{K}{}^2
	=  	  \kappa^2(\partial_y \overset{(0)}{\varphi} )^2
	  + {12 \over \ell^2 } \ ,  \\
    	\label{sodeq:hamiltonian0} 
  && \partial^2_y \overset{(0)}{\varphi} 
        - \overset{(0)}{K} \partial_y  \overset{(0)}{\varphi} =0  \ .
	\label{sodeq:scalar0}	
\end{eqnarray}
The  junction condition (\ref{sodJC:scalar}) at this order
$
  \Bigl[\partial_y \overset{(0)}{\varphi} \Bigr]\bigg|_{y=0} = 0
$
tells us that the solution of Eq.(\ref{sodeq:scalar0}) must be
$
\overset{(0)}{\varphi}=\eta~(x^\mu) \ ,
$
where $\eta (x^\mu )$ is an arbitrary constant of integration.
 Now, the solution of  Eq.(\ref{sodeq:hamiltonian0}) yields
$
\overset{(0)}{K}= {4 / \ell} \ .
$
Other Eqs. (\ref{sodeq:momentum}) and (\ref{sodeq:evolution})
 are trivially satisfied at zeroth order. 
Using the definition
$\overset{(0)}{K}{}_{\mu\nu}= -\overset{(0)}{g}_{\mu\nu,y}/2$,  
we have the lowest order metric 
\begin{equation}
\overset{(0)}{g}{}_{\mu\nu} (y,x^\mu ) 
  = b^2(y)~h_{\mu\nu} (x^\mu ), \qquad 
b(y)\equiv e^{- y/ \ell} \ ,
\label{sodformal:metric}
\end{equation}
where the induced metric on the brane, $ h_{\mu\nu}\equiv g_{\mu\nu} 
(y=0 ,x^\mu) $, arises as a constant of  integration. 
The junction condition for the induced metric (\ref{sodJC:metric0}) merely 
implies well known relation $\kappa^2\sigma =6/\ell$
and that for the scalar field (\ref{sodJC:scalar}) is trivially satisfied.
At this leading order analysis, we can not determine the constants of 
integration $h_{\mu\nu} (x^\mu)$ and $\eta (x^\mu)$ which are constant
 as far as the short length scale $\ell$ variations are concerned, but
 are allowed to vary over the long wavelength scale. These constants
 should be  constrained by the next order analysis. 

 Now, we take into account the effect of both the bulk scalar field 
and the matter on the brane perturbatively.
Our iteration scheme 
is to write the metric $g_{\mu\nu}$ and the scalar field $\varphi$ 
as a sum of local tensors built out of the  induced metric and the induced 
 scalar field on the brane,  in the order of expansion parameters, that is, 
$O((R \ell^2 )^{n})$ and $O(\kappa^2\ell^2V(\varphi))^n$, 
$n=0,1,2,\cdots$~\cite{sodKS4}. 
 Then, we expand the metric and the scalar field  as 
\begin{eqnarray}
&& g_{\mu\nu} (y,x^\mu ) =
	b^2 (y) \left[ h_{\mu\nu} (x^\mu) 
  	+ \overset{(1)}{g}_{\mu\nu} (y,x^\mu)
      	+ \overset{(2)}{g}_{\mu\nu} (y, x^\mu ) + \cdots  \right]  \ , 
      	\label{expansion:metric} \nonumber\\
&& \varphi(y,x^\mu) = \eta (x^\mu) 
	+ \overset{(1)}{\varphi} (y, x^\mu) +\overset{(2)}{\varphi} (y, x^\mu)
	+ \cdots  \ .
\end{eqnarray}
Here, we put the boundary conditions
 $ \overset{(i)}{g}_{\mu\nu} (y=0 ,x^\mu ) =  0  
	\ ,   \overset{(i)}{\varphi } (y=0 ,x^\mu) =0 \ ,\quad i=1,2,3,...
$
so that we can interpret $h_{\mu\nu}$ and $\eta$ as induced quantities. 
Extrinsic curvatures can be also expanded as 
\begin{eqnarray}
K = \frac{4}{\ell} 
	+\overset{(1)}{K}
	+\overset{(2)}{K}
	+\cdots  \ , \qquad
\Sigma^\mu_{\ \nu} = 
	\overset{(1)}{\Sigma}{}^{\mu}_{\ \nu}
        +\overset{(2)}{\Sigma}{}^{\mu}_{\ \nu}
        +\cdots \ .
        \label{expansion:sigma}
\end{eqnarray}
 Using the formula such as 
$\overset{(4)}{R} (\overset{(0)}{g}_{\mu\nu})=R(h_{\mu\nu})/b^2$, 
we obtain the solution
\begin{eqnarray}
\overset{(1)}{K}=\frac{\ell}{6b^2}\left(
	R (h) -\kappa^2\eta^{|\alpha}\eta_{|\alpha}\right)
	-\frac{\ell}{3}\kappa^2V(\eta) \ ,
	\label{sod1:hamiltonian} 
\end{eqnarray}
where $R(h)$ is the scalar curvature of $h_{\mu\nu}$
 and $|$ denotes the covariant derivative with respect to  
$h_{\mu\nu}$. 
 Substituting the results at zeroth order solutions into 
Eq.~(\ref{sodeq:evolution}), we obtain
\begin{eqnarray}
\overset{(1)}{\Sigma}{}^{\mu}{}_{\nu}=
	\frac{\ell}{2b^2}\left[
	R^{\mu}{}_\nu (h) -\kappa^2\eta^{|\mu}\eta_{|\nu}\right]_{\rm traceless}
	+\frac{\chi^{\mu}{}_\nu}{b^4} \ ,
	\label{sod1:evolution}
\end{eqnarray}
where $R^{\mu}{}_\nu (h)$ denotes the Ricci tensor of $h_{\mu\nu}$ and 
$\chi^{\mu}{}_\nu$ is a constant of integration which 
satisfies the constraint $\chi^{\mu}{}_{\mu}=0$. 
 Hereafter, we omit the argument of the curvature for simplicity. 
 Integrating the scalar field equation (\ref{sodeq:scalar}) at first order, 
we have
\begin{eqnarray}
\partial_y\overset{(1)}{\varphi}&=&\frac{\ell}{2b^2} \Box \eta
	-\frac{\ell}{4}V'(\eta)
	+\frac{C}{b^4} \ ,
	\label{sod1:scalar}
\end{eqnarray}
where $C$ is also a constant of integration. 
At first order in this iteration scheme,
we get two kinds of constants of integration, $\chi^{\mu}{}_{\nu}$ and $C$.

Given  the matter fields $T_{\mu\nu}$ on the brane,  the junction 
condition (\ref{sodJC:metric}) becomes
\begin{eqnarray}
\left[\overset{(1)}{\Sigma}{}^{\mu}{}_{\nu}
	-\frac{3}{4}\delta^\mu_\nu
	\overset{(1)}{K}\right] \Bigg|_{y=0} 
	=    \frac{\kappa^2}{2}T^{\mu}{}_{\nu} \ .
	\label{sod1:JC-m}
\end{eqnarray}
 At this order, the junction condition (\ref{sodJC:scalar}) yields
\begin{eqnarray}
\biggl[\partial_y\overset{(1)}{\varphi}\biggr] \Bigg|_{y=0}
	=  0  \ .
	\label{sod1:JC-s}
\end{eqnarray}
These junction conditions give the effective equations of motion
 on the brane.

 Now, we are in a position to discuss the effective equations of motion
 for the dilatonic two-brane models.
The point is the fact that the equations of motion on each brane
take the same form if we use the induced metric on each brane~\cite{sodKS4}. 
 The effective Einstein equations on each positive ($\oplus$)
and negative ($\ominus$) tension brane at low-energies yield
\begin{eqnarray}
G^\mu{}_\nu (h) &=& \kappa^2 \left( 
	\eta^{|\mu}\eta_{|\nu}
        -{1\over 2}\delta^\mu_\nu \eta^{|\alpha} \eta_{|\alpha}
        -{1\over 2}\delta^\mu_\nu V\right) 
        -{2\over \ell} \chi^\mu{}_\nu
        +\frac{\kappa^2}{\ell}  
        \overset{\oplus}{T}{}^{\mu}{}_{\nu}
        \label{sodp:einstein}    \ , \\
G^\mu{}_\nu (f) &=& \kappa^2 \left( 
        \eta^{;\mu} \eta_{;\nu}
        -{1\over 2}\delta^\mu_\nu \eta^{;\alpha} \eta_{;\alpha}
        -{1\over 2}\delta^\mu_\nu V\right)
        -{2\over \ell} {\chi^\mu{}_\nu \over \Omega^4}
        -\frac{\kappa^2}{\ell} 
        \overset{\ominus}{T}{}^{\mu}{}_{\nu} 
        \label{sodn:einstein-1}  \ .
\end{eqnarray}
where $f_{\mu\nu}$ is the induced metric on the negative tension brane 
and $;$ denotes the covariant derivative with respect to $f_{\mu\nu}$. 
 When we set the position of the positive tension brane at $y=0$, 
  that of the negative tension brane $\bar{y}$ in general depends on
 $x^\mu$, i.e. $\bar{y} = \bar{y} (x^\mu)$. Hence,  
 the warp factor at the negative tension brane 
 $\Omega (x^\mu)\equiv b(\bar{y}(x)) $ also depends on $x^\mu$.
  Because the metric always comes into equations
 with  derivatives, the zeroth order relation is enough in this 
 first order discussion. Hence, the metric on the
 positive tension brane is related to the metric on the 
 negative tension brane as $f_{\mu\nu} = \Omega^2 h_{\mu\nu}$.
   
Although Eqs.~(\ref{sodp:einstein}) and (\ref{sodn:einstein-1}) are non-local 
individually, with undetermined $\chi^{\mu}{}_{\nu}$, one can combine both
equations to reduce them to local equations for each brane. 
We can therefore easily eliminate $\chi^{\mu}{}_{\nu}$ from 
Eqs.~(\ref{sodp:einstein}) and (\ref{sodn:einstein-1}), since 
$\chi^{\mu}{}_{\nu}$ appears only algebraically.
Eliminating $\chi^\mu{}_\nu$ from both Eqs.~(\ref{sodp:einstein}) and 
(\ref{sodn:einstein-1}), we obtain 
\begin{eqnarray}
G^\mu{}_\nu &=&
	\frac{\kappa^2}{\ell\Psi}
	\overset{\oplus}{T}{}^{\mu}{}_{\nu}
	+\frac{\kappa^2(1-\Psi)^2}{\ell\Psi}
	\overset{\ominus}{T}{}^{\mu}{}_{\nu}
        \nonumber\\
&&
	+{1\over \Psi}\left[ \Psi^{|\mu}{}_{|\nu} 
     	- \delta^\mu_\nu \Psi^{|\alpha}{}_{|\alpha} 
     	+{3\over 2}{1 \over 1-\Psi} \left(
     	\Psi^{|\mu} \Psi_{|\nu} -{1\over 2}\delta^\mu_\nu
     	\Psi^{|\alpha} \Psi_{|\alpha} \right) \right] \nonumber\\
&& 
	+\kappa^2\left( 
        \eta^{|\mu} \eta_{|\nu}
        -\frac{1}{2}\delta^\mu_\nu\eta^{|\alpha}\eta_{|\alpha}
	-\delta^\mu_\nu V_{\rm eff}
        \right)
        \label{sodQST1}
         \ ,  \quad V_{\rm eff}=  \frac{2-\Psi}{2}V \ ,
\end{eqnarray}
where we defined a new field $\Psi=1-\Omega^2$ 
which we refer to by the name ``radion".
 The bulk scalar field induces the energy-momentum
 tensor of the conventional 4-dimensional scalar field with 
 the effective potential which depends on the radion. 

We can also determine the dark radiation $\chi^{\mu}{}_{\nu}$ by eliminating 
$G^{\mu}{}_{\nu}(h)$ from Eqs.~(\ref{sodp:einstein}) and
(\ref{sodn:einstein-1}),
\begin{eqnarray}
{2\over \ell}\chi^\mu{}_{\nu}
	&=&-{1\over \Psi}\left[ \Psi^{|\mu}{}_{|\nu} 
     	- \delta^\mu_\nu \Psi^{|\alpha}{}_{|\alpha} 
     	+{3\over 2}{1 \over 1-\Psi} \left(
     	\Psi^{|\mu} \Psi_{|\nu} -{1\over 2}\delta^\mu_\nu
     	\Psi^{|\alpha} \Psi_{|\alpha} \right) \right] 
     	\nonumber\\
&&
     	+{\kappa^2 \over 2}(1-\Psi)\delta^\mu_\nu V
     	-\frac{\kappa^2}{\ell}\frac{1-\Psi}{\Psi}
     	\left[\overset{\oplus}{T}{}^{\mu}{}_{\nu}
     	+  \left(1-\Psi\right)
     	\overset{\ominus}{T}{}^{\mu}{}_{\nu}   \right] \ .
     	\label{sodchi2}
\end{eqnarray}
Due to the property $\chi^\mu{}_\mu =0$, we have
\begin{eqnarray}
\Box\Psi =
	\frac{\kappa^2}{3\ell}(1-\Psi)\left[
  	\overset{\oplus}{T}
  	+(1-\Psi)\overset{\ominus}{T}\right]
	-{1\over 2(1-\Psi)} \Psi^{|\alpha} \Psi_{|\alpha} 
  	-{2\kappa^2\over 3}\Psi(1-\Psi)V   
  	\label{sodQST2}\ .
\end{eqnarray}
Note that Eqs.~(\ref{sodQST1}) and (\ref{sodQST2}) are derived from 
a scalar-tensor type theory coupled to the  additional scalar field.

Similarly, the equations for the scalar field on branes become
\begin{eqnarray}
&&\Box_h\eta-{V^\prime\over 2}+{2\over\ell}C
	=0  \ ,
	\label{sodp:KG}\\
&&\Box_f\eta-{V^\prime\over 2}+{2\over\ell}{C\over\Omega^4}
	=0 \ ,
	\label{sodn:KG}
\end{eqnarray}
where the subscripts refer to the induced metric on each brane. 
Notice that the scalar field takes the same value for both branes 
at this order.
Eliminating the dark source $C$ from these Eqs.~(\ref{sodp:KG}) 
and (\ref{sodn:KG}), we find the equation for the scalar field 
takes the form
\begin{eqnarray}
\Box_h\eta-V'_{\rm eff}
	=-\frac{\Psi^{|\mu}}{\Psi}\eta_{|\mu} \ .
	\label{sodp:KG-2}
\end{eqnarray}
Notice that the radion acts as a source for $\eta$. And we can also 
get the dark source as
\begin{eqnarray}
{2\over \ell}C=-{V^\prime\over 2}(1-\Psi)
	+{\Psi^{|\mu}\over\Psi}\eta_{|\mu}    \ .
	\label{sodC2}
\end{eqnarray}

Now the effective action for the positive tension brane which
gives Eqs.~(\ref{sodQST1}), (\ref{sodQST2}) and (\ref{sodp:KG-2}) can be read off as
\begin{eqnarray}
S&=&{\ell\over 2\kappa^2}\int d^4x\sqrt{-h}\left[
	\Psi R
	-\frac{3}{2(1-\Psi)}\Psi^{|\alpha} \Psi_{|\alpha} 
	-\kappa^2\Psi\left(\eta^{|\alpha}\eta_{|\alpha}
	+2V_{\rm eff} (\eta,\Psi) \right)
	\right]  
	\nonumber\\
&&
	+\int d^4x\sqrt{-h}~
	\overset{\oplus}{\cal L}
	+\int d^4x\sqrt{-h}~(1-\Psi)^2
	\overset{\ominus}{\cal L}  \ ,
\end{eqnarray}
where the last two terms represent actions for the matter on each brane.  
Thus, we found the radion field couples with the induced metric and 
the induced scalar field on the brane non-trivially. 
 Surprisingly, at this order, the nonlocality of 
 $\chi_{\mu\nu}$ and $C$
 are eliminated by the radion. 
We see the radion has a conformal coupling. However, in the present case,
the radion couples to the dilaton field which breaks a conformal invariance.
Hence, this gives non-conformal holography. 

 As this is a closed system, we can analyze a primordial spectrum
 to predict the  cosmic background fluctuation spectrum~\cite{sodSoda:2005mu}.
  Interestingly, $\chi^{\mu}_{\nu}$ and $C$ vanishes in the single brane 
limit, $\Psi \rightarrow 1$, as can be seen from (\ref{sodchi2}) and (\ref{sodC2}).
 The dynamics is simply governed by Einstein theory with the single scalar 
 field. Therefore, we can conclude that the bulk inflaton can drive inflation
 when the slow role conditions are satisfied.

\subsection{AdS/CFT and KK corrections: Single-brane cases }

 It would be important to take into account the KK effects
 as corrections to the leading order result. 
 It can be accomplished in the single-brane models. 
 Using our approach, in the single brane limit,
 we can deduce the effective action  with KK corrections as~\cite{sodKS4,sodKanno:2003xy}
 ( see also  \cite{sodSoda:2002ky,sodSoda:2004hq} )
\begin{eqnarray}
S&=&\frac{\ell}{2\kappa^2}\int d^4x\sqrt{-h}\left[
	\left(1+\frac{\ell^2}{12}\kappa^2V\right)R
	-\kappa^2\left(
	1+\frac{\ell^2}{12}\kappa^2V-\frac{\ell^2}{4}V''\right)
	\eta^{|\alpha}\eta_{|\alpha}
	-2\kappa^2V_{\rm eff}
	\right.\nonumber\\
&&\left.\hspace{2.5cm}
	-\frac{\ell^2}{4}\left(
	R^{\alpha\beta}R_{\alpha\beta}
	-\frac{1}{3}R^2
	\right)\right]
	+\int d^4 x \sqrt{-h}~ 
	{\cal L}_{\rm matter}
	+S_{\rm CFT} \ ,
\end{eqnarray}
where the last term comes from 
 the energy-momentum tensor of CFT matter $\tau_{\mu\nu}$
 and the effective potential at this order is defined by 
\begin{eqnarray}
V_{\rm eff}=\frac{1}{2}V
	+\frac{\ell^2\kappa^2}{48}V^2
	-\frac{\ell^2}{64}V^{\prime 2} 
	\label{2:effpt}\ .
\end{eqnarray}
 It is interesting to note that the effective potential contains the terms
 which looks like F-terms in supersymmetric models.

Thus, even in the dilatonic braneworld, the AdS/CFT correspondence seems to play
an important role in the single-brane case.

\section{Conclusion}

 In this lecture, I have reviewed the gradient expansion method
 in the context of braneworlds. 
 Using the formalism, I have
 tried to explain how the AdS/CFT correspondence is
 related to the braneworld models. 
 
 In the case of the RS single-brane model, we clarified when the conventional
 Einstein equations hold at low energy. 
 Moreover, we revealed the relation between
   the geometrical and the AdS/CFT correspondence 
   approach using the gradient expansion method.
We have shown that the high energy and the Weyl term corrections found in the geometrical
 approach correspond to the CFT matter corrections found in the 
    AdS/CFT approach.   
    
 In the case of the RS two-brane sysytem, 
    we showed that the AdS/CFT correspondence plays
    an important role in the sense that the low energy effective field theory
    can be described by the conformally coupled scalar-tensor theory
     where the radion plays the role of the scalar field.
 We also presented the symmetry method to derive KK corrections in the two-brane
 system.
 
These effective theories for RS braneworlds can be used to make cosmological predictions.
 More importantly, it turned out that the gradient expansion method provides
 a unified view of RS braneworlds. 

 We have also considered the bulk scalar field  with a nontrivial potential
 and derived the non-linear low energy effective action for the
 dilatonic two-brane model using the gradient expansion method.
 As a result, we have shown that the effective theory reduces to the scalar-tensor
 theory with the non-trivial coupling between the radion and the bulk scalar
 field.  Since the radion has a conformal coupling, the conformal
 symmetry is relevant even for the dilatonic braneworlds.
 In this sense, the AdS/CFT correspondence is related to dilatonic braneworlds.
 However, the radion couples to the scalar field which is non-conformal.
 Hence, the conformal invariance is violated. 

 Our phenomenological motivation to consider dilatonic braneworlds
 was a possibility of the bulk inflaton. Concerning to this issue,
 taking into account the fact that $\chi_{\mu\nu}$ and $C$ becomes zero 
 when two branes get separated  infinitely,
 one can conclude that the bulk inflaton can drive the inflation on the brane
 as far as the slow roll conditions are satisfied. We also obtained
 KK corrections in the single brane limit which contain CFT corrections. 

 These results tell us that there exist profound relations between braneworlds
 and the AdS/CFT correspondence, although the correspondence is slightly
 deformed in the dilatonic cases. 

In this lecture, we have considered only codimesion-one braneworlds.
It is important to extend the analysis to higher codimension
 models~\cite{sodCline:2003ak,sodBurgess:2004dh,sodVinet:2005dg,sodBostock:2003cv,sodKanno:2004nr,sodCharmousis:2009uk,sodPapantonopoulos:2007fk,
 sodPapantonopoulos:2006uj,sodKanno:2007wj,sodChen:2008sn}.
It is intriguing to study a role of the AdS/CFT correspondence
in these higher codimension braneworlds.

\begin{acknowledgement}
I would like to thank Sugumi Kanno 
for collaborations on which this lecture
is based and useful comments on the manuscript.
I am grateful to the organizers of the 5th Aegean Summer school, especially 
Lefteris Papantonopoulos, for inviting me to give a lecture and their
kind hospitality during the summer school. The present work
 is supported by the Japan-U.K. Research Cooperative Program, 
Grant-in-Aid for  Scientific Research Fund of the Ministry of 
Education, Science and Culture of Japan No.18540262,
Grant-in-Aid for  Scientific Research on Innovative Area No.21111006
and the Grant-in-Aid for the Global COE Program 
``The Next Generation of Physics, Spun from Universality and Emergence".
\end{acknowledgement}

%
%
%

\end{document}